# ComPPI: a cellular compartment-specific database for protein–protein interaction network analysis

Daniel V. Veres[1,†], Dávid M. Gyurkó[1,†], Benedek Thaler[1,2], Kristóf Z. Szalay[1], Dávid Fazekas[3], Tamás Korcsmáros[3,4,5] and Peter Csermely[1,*]

[1]Department of Medical Chemistry, Semmelweis University, Budapest, Hungary, [2]Faculty of Electrical Engineering and Informatics, Budapest University of Technology and Economics, Budapest, Hungary, [3]Department of Genetics, Eötvös Loránd University, Budapest, Hungary, [4]TGAC, The Genome Analysis Centre, Norwich, UK and [5]Gut Health and Food Safety Programme, Institute of Food Research, Norwich, UK

**ABSTRACT**

Here we present ComPPI, a cellular compartment-specific database of proteins and their interactions enabling an extensive, compartmentalized protein–protein interaction network analysis (URL: http://ComPPI.LinkGroup.hu). ComPPI enables the user to filter biologically unlikely interactions, where the two interacting proteins have no common subcellular localizations and to predict novel properties, such as compartment-specific biological functions. ComPPI is an integrated database covering four species (*S. cerevisiae, C. elegans, D. melanogaster* and *H. sapiens*). The compilation of nine protein–protein interaction and eight subcellular localization data sets had four curation steps including a manually built, comprehensive hierarchical structure of >1600 subcellular localizations. ComPPI provides confidence scores for protein subcellular localizations and protein–protein interactions. ComPPI has user-friendly search options for individual proteins giving their subcellular localization, their interactions and the likelihood of their interactions considering the subcellular localization of their interacting partners. Download options of search results, whole-proteomes, organelle-specific interactomes and subcellular localization data are available on its website. Due to its novel features, ComPPI is useful for the analysis of experimental results in biochemistry and molecular biology, as well as for proteome-wide studies in bioinformatics and network science helping cellular biology, medicine and drug design.

## INTRODUCTION

Biological processes are separated in the cellular and subcellular space, which helps their precise regulation. Compartmentalization of signalling pathways is a key regulator of several main biochemical processes, such as the nuclear translocation-mediated activation of transcription factors (1). Several proteins are located in more than one subcellular localizations. As an example, IGFBP-2 is a predominantly extracellular protein with a key role in insulin growth factor signalling (2), while its translocation into the nucleus results in vascular endothelial growth factor-mediated angiogenesis (3). Another important example is the HIF-1 Alpha with translocation from the cytosol to the nucleus, where it acts as a transcription factor involved in the maintenance of cellular oxygen homeostasis (4) (Supplementary Figure S1). Their shuttling between these localizations is a key regulatory mechanism, which implicates the importance of improving the systems level analysis of compartmentalized biological processes.

Protein–protein interaction data are one of the most valuable sources for proteome-wide analysis (5), especially to understand human diseases on the systems-level (6) and to help network-related drug design (7). However, protein–protein interaction databases often contain data with low overlap (8), and are designed using different protocols (9), therefore, their integration is needed to improve our comprehensive knowledge (10). Low-throughput data sets often use several different protein naming conventions causing difficulties in data analysis and integration. Manual curation of data yields a large improvement of data quality (11).

Interaction data often contain interactions, where the two interacting proteins have no common subcellular localizations (12). These interactions could be biophysically possible, but biologically unlikely (13). Thus, these interactions cause data bias that leads to deteriorated reliability in interactome-based studies (14), especially those involv-ing subcellular localization-specific cellular processes (15). Unfortunately, subcellular localization data are incomplete. Despite the need of experimentally verified subcellular localizations for reliable compartmentalization-based interactome filtering (16), only computationally predicted subcellular localization information is available for a large part of the proteome. Moreover, subcellular localization data are redundant, often poorly structured and miss to highlight the reliability of data (17).

*To whom correspondence should be addressed. Tel: +361 459 1500; Fax: +361 266 3802; Email: csermely.peter@med.semmelweis-univ.hu
†The authors wish it to be known that, in their opinion, the first two authors should be regarded as Joint First Authors.

Existing analysis tools involving subcellular localizations offer the download of filtered interactomes for a subset of proteins (like MatrixDB (18)). Several databases use only Gene Ontology (GO (19)) cellular component terms as the source of the subcellular localization data (such as HitPredict (20) or Cytoscape BiNGO plugin (21)), while GO still contains data inconsistency despite its highly structured annotations (22). Cytoscape Cerebral plugin (23) generates a view of the interactome separated into layers according to their subcellular localization. In different data sets the subcellular localization structure is not uniform, which makes their comparisons often difficult.

ComPPI-based interactomes introduced here provide a broader coverage (Supplementary Tables S1 and S2), using several curation steps in data integration. ComPPI offers highly structured subcellular localization data supplemented with Localization and Interaction confidence Scores, all presented with user-friendly options. As a key feature ComPPI allows the construction of high-confidence data sets, where potentially biologically unlikely interactions in which the interacting partners are not localized in the same cellular compartment, have been deleted. As our examples will show, this gives novel options of interactome analysis and also suggests potentially new subcellular localizations and localization-based functions.

## DESCRIPTION OF THE DATABASE

### Overview of ComPPI

Our goal by constructing ComPPI was to provide a reliable subcellular compartment-based protein–protein interaction database for the analysis of biological processes on the subcellular level. A key feature of ComPPI is that it allows the filtering of localization-based biologically unlikely interactions resulting in localization-wise more reliable interaction data. During the integration of 17 databases to build up ComPPI, we used the following four curation steps to improve data quality (Figure 1). (i) Source databases were selected by comparing them to a large number of other potential databases and their data content was manually reviewed. (ii) Subcellular localization data were consistently structured to a hierarchical subcellular localization tree (Supplementary Figure S2) containing more than 1600 individual sublocalizations. (iii) We developed an algorithm to map different protein naming conventions to UniProt accession numbers (24,25). (iv) Finally, a manual follow-up by six independent experts was performed in order to revise the data content searching for data inconsistence and false entries, and to test the functions of the web interface (Supplementary Table S3).

ComPPI database includes comprehensive and integrated data of four species (*Saccharomyces cerevisiae*,

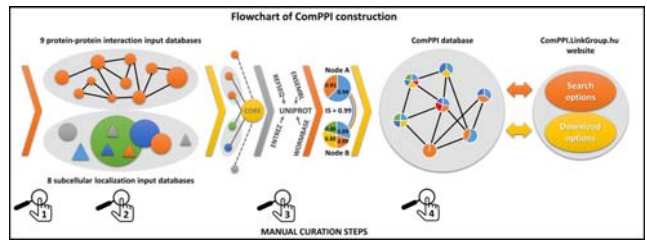

**Figure 1.** Flowchart of ComPPI construction highlighting the four curation steps. Constructing the ComPPI database we first checked the data content of 24 possible input databases for false entries, data inconsistence and compatible data structure in order to minimize the bias in ComPPI coming from the input sources **(1)**. As a consequence we selected nine protein–protein interaction (BioGRID (29), CCSB (30), DiP (31), DroID (26), HPRD (27), IntAct (32), MatrixDB (18), MINT (33) and MIPS (28)) and eight subcellular localization databases (eSLDB (37), GO (19), Human Proteinpedia (34), LOCATE (38), MatrixDB (18), OrganelleDB (39), PA-GOSUB (36) and The Human Protein Atlas (35)) in order to integrate them into the ComPPI data set. The subcellular localization structure was manually annotated creating a hierarchic, non-redundant subcellular localization tree using >1600 GO cellular component terms (19) for the standardization of the different data resolution and naming conventions **(2)**. All input databases were connected to the ComPPI core database with newly built interfaces in order to improve data consistency, to allow easy extensibility with new databases and to incorporate automatic database updates. As part of the curation steps the filtering efficiency of our newly built interfaces were tested on 200 random proteins for every input databases, and the interfaces were accepted only when all the requested false-entries and data content errors were filtered, in order to establish a more reliable content (Supplementary Table S3). During data integration, different protein naming conventions were mapped to the most reliable protein name. In this process we used publicly available mapping tables (UniProt (24) and HPRD (27)). For 30% of protein names we applied manually built mapping tables with the help of online ID cross-reference services (PICR (25) and Synergizer (http://llama.mshri.on.ca/synergizer/translate/)) **(3)**. After data integration Localization and Interaction Scores were calculated (for detailed description see Figure 2). As an illustration we show the example of Figure 2 with two interacting proteins (nodes A and B corresponding to HSP 90-alpha A2 and Survivin, respectively) with shared cytosolic and nuclear localizations (light blue and orange). Node B has an additional membrane (yellow) subcellular localization and an extracellular localization (green). Numbers in the circles of nodes A and B refer to their Localization Scores. The Interaction Score of nodes A and B is 0.99 (see Figure 2 for details). The integrated ComPPI data set was manually revised by six independent experts **(4)**. During the revision two of the six experts tested our database on 200 random proteins each to ensure high-quality control requirements, and searched for exact matches between the entries in the input sources and the ComPPI data set. All the experts searched for false entries, data inconsistency, protein name mapping errors in the downloadable data and tested the operation of the online services as well. After the revision we updated our source databases, their interfaces, the subcellular localization tree and the algorithm generating the downloadable data, in order to acquire all the changes proposed during the tests. As the final result, the webpage http://ComPPI.LinkGroup.hu is available for search and download options in order to extract the biological information in a user-friendly way.

*Caenorhabditis elegans*, *Drosophila melanogaster* and *Homo sapiens*) cataloguing 125 757 proteins, their 791 059 interactions and 195 815 major subcellular localizations in its current, 1.1 version. The proteome-wide data set contains localizations for five main subcellular organelles (nucleus, mitochondrion, cytosol, secretory-pathway, membrane) and the extracellular compartment. Importantly, 60% of the ComPPI entries have high resolution cellular localization data assigning them to one or several of >1600 GO cellular component terms (19) associating these proteins with distinct subcellular compartments.

**Design and implementation**

Both protein–protein interaction and subcellular localization data are incorporated to ComPPI automatically using their own interface to bridge the difference in data structure (Supplementary Table S3). New interfaces can be added without limitations. The incoming data are merged to form a consistent internal data pool using a comprehensive protein name mapping algorithm, in order to deal with the redundancy in the input data sets (http://comppi.linkgroup.hu/help/naming_conventions). The website and the downloadable contents are generated from this integrated internal data pool. All curated parts are stored in separate, yet interconnected containers to maintain persistency between ComPPI releases.

The website follows the hierarchical model-view-controller design pattern to ensure the separation of the data layer from the business logic and the user interface. Each functional unit is implemented as a module to further support easy maintenance and extensibility. Protein search algorithms have been extensively optimized, and the served content is cached to ensure quick response times even on low-end infrastructure. Due to these features ComPPI can be easily run on a general laptop or desktop computer.

The downloadable data sets are pre-generated and validated automatically and manually in every release to fulfill our high quality control requirements (Figure 1). The Python script that generates these data sets also contains basic tools for data retrieval and manipulation in a network-oriented manner, which enables the user to perform bioinformatics analysis on the interactome using the open source code and also gives space for further improvement.

End-user documentation is available at the website as tutorials, detailed descriptions and location-specific tooltips. All components of ComPPI and the underlying software stack are open source. The source code is available in a revision controlled repository at http://bificomp2.sote.hu:22422/comppi/summary.

Third-party tools and technologies were selected with open accessibility and scientific reproducibility in mind including the Ubuntu Linux 14.04 operating system (http://ubuntu.com/), the nginx HTTP server (http://nginx.org/), the MySQL 5 Community Edition database server (http://www.mysql.com/), the git version control system (http://git-scm.com/), the PHP 5 scripting language (https://php.net/), the Symfony 2 PHP framework (http://symfony.com/), the jQuery JavaScript framework (http://jquery.com/), the D3.js JavaScript library for network visualization (http://d3js.org/) and the Python3 scripting language (https://python.org/).

**Database content and access**

*Input databases.* The low overlap of protein–protein interaction and subcellular localization databases (11) prompted us to integrate several source databases in order to improve data coverage and quality (Supplementary Figure S3 and Supplementary Table S2). In this process we used publicly downloadable license-free data sources, preferably containing proteome-wide data sets. Protein–protein interaction data were selected to contain only physical interactions with experimental evidence coming from high-throughput, as well as low-throughput techniques. We incorporated the widely used species-specific (DroID (26), HPRD (27), MatrixDB (18) and MIPS (28)) and general (BioGRID (29), CCSB (30), DiP (31), IntAct (32) and MINT (33)) protein–protein interaction databases having high data quality, update frequency and freely downloadable latest releases for academic research.

Subcellular localization data can be obtained from experimental evidence or using predictions. Several source databases contained only experimentally verified subcellular localization entries (such as Human Proteinpedia (34) and the Human Protein Atlas (HPA) (35)). Other source data had only computationally predicted information (such as PA-GOSUB (36)). Several data sources had integrated data structure (such as eSLDB (37), GO (19), LOCATE (38), MatrixDB (18), OrganelleDB (39)) containing data of both experimental and predicted origin. During the selection of the proteome-wide predicted subcellular localization databases with downloadable content we focused on the use of prediction algorithms with combined methods using robust machine learning tools validated on highly reliable training sets.

*ComPPI data set.* The availability of the data sources differs between various species. As an example ComPPI contains eight protein–protein interaction and eight subcellular localization databases for human proteins (Supplementary Figure S3). Database integration was based on protein ID mapping to the most reliable naming convention available, primarily to UniProt Swiss-Prot accession numbers (11). The 4 curation steps (Figure 1) allow the users to access interaction and localization data at a single resource having a higher coverage and reliability than the incorporated databases.

The ComPPI database contains three types of predefined data sets: (i) the compartmentalized interactome catalogues of those protein–protein interactions, where the interacting proteins have at least one common subcellular localization, (ii) the integrated protein–protein interaction data set which can be customized by the four species included and (iii) the subcellular localization data set, which is one of the biggest existing subcellular localization resource with a comprehensive structure for interactome analysis. All downloadable ComPPI resources are license free and publicly available for academic and industrial research.

**Search and download features**

*Search features.* The internally hyper-linked web application of ComPPI enables even those users, who have no bioinformatics expertise, to search for the interactions of individual proteins. Search options (http://comppi.linkgroup.hu/protein_search) are available for protein names with autocomplete function giving their subcellular localization, their interactions and the likelihood of their interactions considering the subcellular localization of the interacting partners. Using the Advanced Settings of the Search page the user is able to filter the list of the possible query proteins for species, subcellular localizations and/or localization probability. These settings can be set for the interactors of the query protein too and are adjustable with the Custom Settings on the Results page, which allows the filtering of the interactors for subcellular localizations, localization probability and interaction score. The properties of the query protein and its interactors are available for download. After filtration only those interactions are exported that fulfill the custom filtering requirements set by the user. Network visualization of the whole or filtered first-neighbour interactome of the query protein is also available, where the width of the edges corresponds to the Interaction Score of the given interaction. These options together provide a user-friendly web interface for data mining for both non-experts and computational biologists. A Direct Search option is also available via URL, which gives the opportunity to interconnect the ComPPI database with other resources, or to generate multiple searches for data mining.

*Download options.* All ComPPI data are available for download at the website. Predefined data sets can be customized by the user to contain only data for a requested species or localization: (i) Compartmentalized interactomes have interactions, where the two interacting protein-nodes have at least one common subcellular localization. These interactomes can be filtered to species besides subcellular localizations. (ii) Integrated protein–protein interaction data sets contain all the interactions, and can be customized to the four species included. (iii) Integrated subcellular localization data sets contain proteins together with their localization data. The user can select species and localizations to customize these data sets. (iv) The current and previous releases of the full database can also be downloaded. A detailed help and a tutorial for the Search and Download functions are both available.

*Output.* ComPPI output data provide lists of interactions, interaction scores of the interacting proteins and localizations with localization scores. Moreover, the user receives the PubMed IDs and references of the source databases for both the interactions and subcellular localizations, and the additional information (if available) of the data type. The user-defined interactomes as results of the Basic or Advanced Search options and the predefined data sets on the Downloads page are available for download in plain text format to ensure convenient data handling. The complete current and previous releases of the database are downloadable in SQL format to provide full access to all the data in ComPPI.

**Localization and interaction scores**

*Subcellular localization structure.* Subcellular localization data are coming from different source databases, containing localizations having experimental evidence (in the followings: experimental), coming from unknown sources (unknown) or predictions (predicted; Figure 2). Experimental data usually have high resolution, where the exact localization of the protein is often defined, such as the nuclear pore complex for Nup107 (40). Predicted localizations have usually low resolution. As an example nuclear localization can be predicted from the existence of a nuclear localization signal in the amino acid sequence (41) without any experimental evidence.

Because of the incongruity in the resolution of the localization data and the different naming conventions between the source databases, we standardized the subcellular localization data using GO cellular component terms (19). In order to solve the problem of the unequivocally mapped GO terms (Figure 3) we created a manually built, non-redundant, hierarchical localization tree (Supplementary Figure S2). With the help of this we clustered the >1600 GO cellular component terms to six major compartments (cytosol, nucleus, mitochondrion, secretory-pathway, membrane, extracellular) (Supplementary Table S4). This new structure allows ComPPI to store all localization entries from different sources and to assign the proteins efficiently to six major compartments (Figure 3 and Supplementary Figure S4).

*Localization and interaction scores.* The ComPPI Localization Score is a novel measure to score the probability of a localization for a given protein. The Localization Score depends on the subcellular localization evidence type (experimental, unknown, predicted) and the number of sources (Figure 2). The Interaction Score characterizes the probability of the subcellular localization of a protein–protein interaction, and is based on the consensus of the compartment-specific Localization Scores of the interacting proteins. With the help of the scoring algorithm ComPPI provides a novel localization probability describing how likely it is that the protein exists in the given subcellular compartment, and gives the opportunity to build high-confidence interactomes based on the distribution of the interaction scores (Supplementary Figure S5).

Localization Scores are calculated using probabilistic disjunction (marked with operator $V$) among the different localization evidence types and the number of ComPPI localization data entries of the respective evidence type (Equation (1), see top panel of Figure 2 for details)

$$\phi_{\text{LocX}} = V_{\text{res}} p_{\text{LocX}} \qquad (1)$$

where $\phi_{\text{LocX}}$ and $p_{\text{LocX}}$ are the Localization Score and the localization evidence type (experimental, unknown or predicted) for protein X and localization Loc, respectively, while res is the number of available ComPPI localization data entries for protein X.

As the first step of Interaction Score calculation, compartment-specific Interaction Scores are obtained by multiplying the Localization Scores of the two interactors for each of the six major compartments. Finally, the Interaction Score is calculated as the probabilistic disjunction (marked with operator $V$) of the Compartment-specific Interaction Scores of all major localizations available for the interacting pair from the maximal number of six major localizations (Equation (2), see bottom panel of Figure 2 for details)

$$\phi_{\text{Int}} = V_{i=1}^{6} \phi_{\text{LocA}} * \phi_{\text{LocB}} \qquad (2)$$

where $\phi_{\text{Int}}$ is the Interaction Score, while $\phi_{\text{LocA}}$ and $\phi_{\text{LocB}}$ are the Compartment-specific Localization Scores of interacting proteins A and B, respectively.

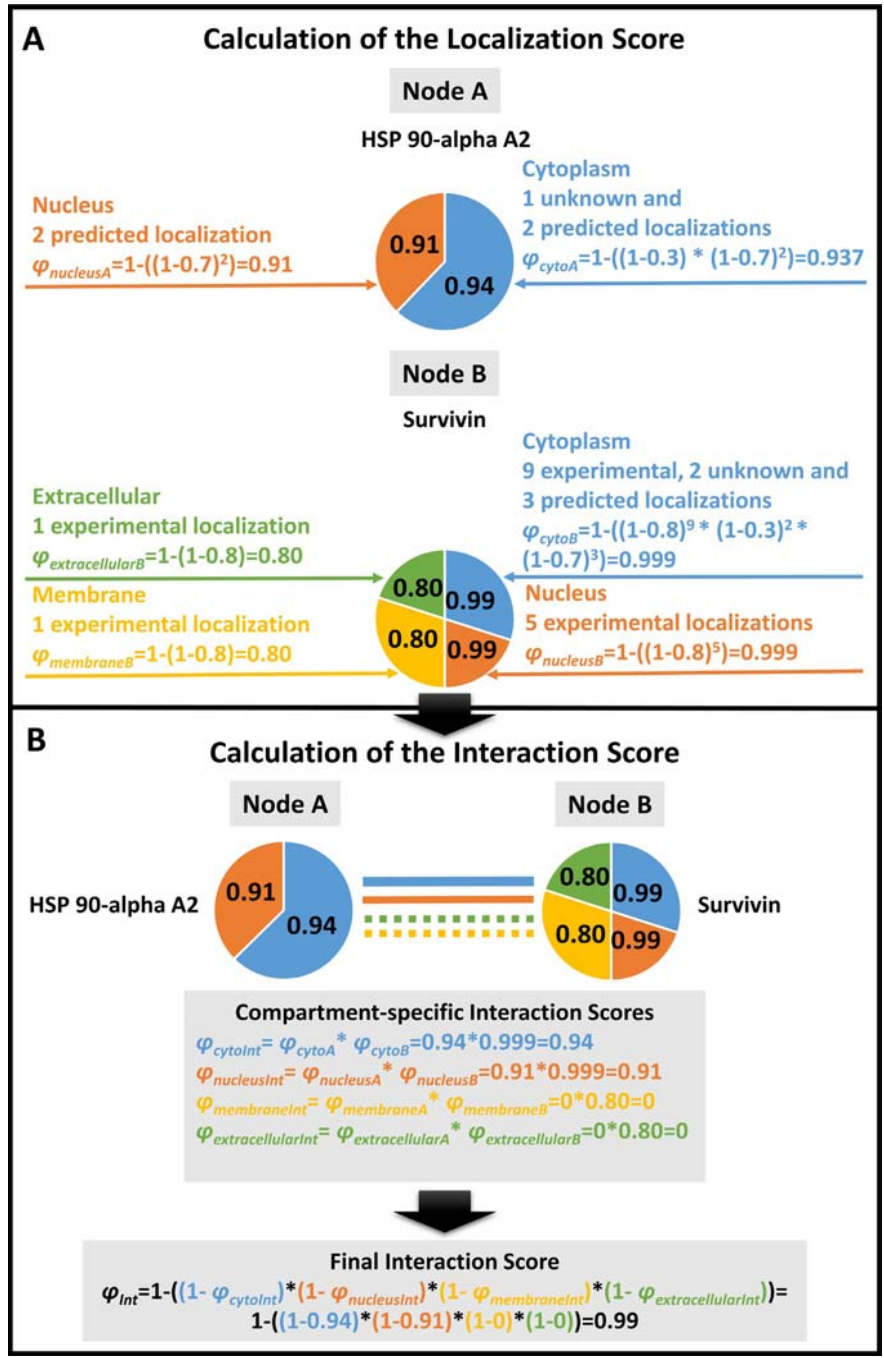

**Figure 2.** Calculation of the subcellular localization-based ComPPI scores. We illustrate the Localization Score calculation steps on the examples of Heat Shock Protein (HSP) 90-apha A2 and Survivin. HSP 90-alpha A2 has two major subcellular localizations, while Survivin has four ($\phi_{nucleusA}$, $\phi_{cytoA}$ and $\phi_{extracellularB}$, $\phi_{membraneB}$, $\phi_{nucleusB}$, $\phi_{cytoB}$, respectively). Localizations were manually categorized into major localizations before the calculation (see the text in section 'Subcellular Localization Structure' for details). (**A**) A Localization Score (such as $<_{cytoA}$) is calculated for every available major subcellular localization for both HSP 90-alpha A2 and Survivin based on the available localization evidence types and the number of the respective localization data entries (corresponding to $p_{LocX}$ and $V_{rec}$ of Equation (1)). The Localization Score calculation uses the optimized localization evidence type weights of 0.8, 0.7 and 0.3 for experimental, predicted or unknown localization evidence types, respectively. (For details of the weight optimization procedure see section 'Score Optimization' of the main text and Supplementary Figure S6.) The Localization Score (i.e. the likelihood for the respective protein to belong to a major compartment) is represented by the probabilistic disjunction among the different localization evidence types and the number of ComPPI localization data entries of the respective evidence type (Equation (1)). (**B**) Calculation of the Interaction Score ($\phi_{Int}$) is based on the Localization Scores of the interacting proteins. First, Compartment-specific Interaction Scores (such as $\phi_{cytoInt}$) are calculated as pair-wise products of the relevant Localization Scores of the two interacting proteins (HSP 90-alpha A2 and Survivin). The final Interaction Score ($\phi_{Int}$) is calculated as the probabilistic disjunction of the Compartment-specific Interaction Scores of all major localizations available for the interacting pair of proteins (in the example four major localizations for HSP 90-alpha A2 and Survivin) from the maximal number of six major localizations (Equation (2)).

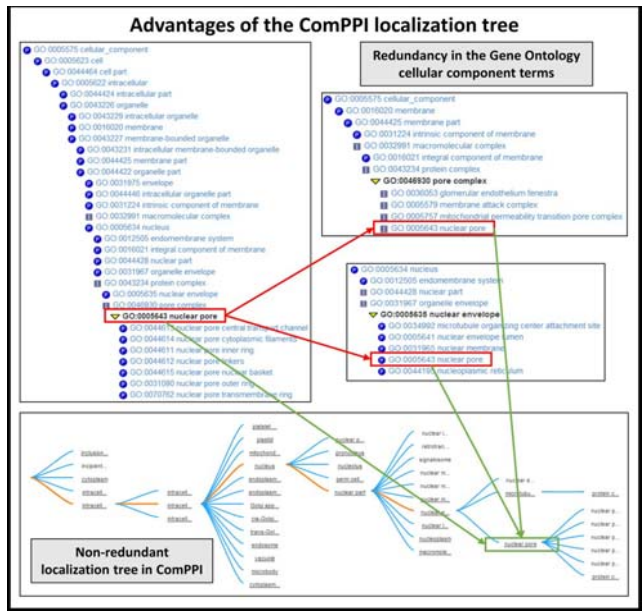

**Figure 3.** Advantages of ComPPI subcellular localization structure. The subcellular localization structure of ComPPI is based on a manually curated, non-redundant subcellular localization tree extracted from GO data (19) containing more than 1600 GO cellular component terms (Supplementary Figure S2). On Figure 3 an example of the redundancy in the GO cellular component tree structure is shown, where the 'nuclear pore' cellular component can be found under several branches in the tree, such as in the 'nucleus' -> 'nuclear envelope' -> 'nuclear pore' or the 'membrane' -> 'membrane part' -> 'intrinsic component of the membrane' -> 'integral component of the membrane' -> 'pore complex' pathways (highlighted in red). Because of the need of the mapping of high-resolution subcellular localization data into major cellular components (Supplementary Table S4) a localization tree with a non-redundant structure was built. In our example, it can be seen that with the help of this structure the 'nuclear pore' derives unequivocally from the 'nuclear envelope' term (highlighted in green).

*Score optimization.* As mentioned before the ComPPI localization evidence type can be experimental, unknown or predicted. ComPPI characterizes each of these localization evidence types by a parameter called the evidence type weight to achieve a unified scoring system applicable to the diverse data sources. To obtain these evidence type weights we performed their data-driven optimization. Based on the fact that experimentally validated entries are the most reliable, while localization entries coming from unknown or predicted origin are less reliable, we set the following order of evidence type weights: experimental > predicted AND experimental > unknown as the two requirements of the optimization process. We chose the HPA database (35) containing only experimentally verified subcellular localizations in order to build a positive control data set, where the interactors have at least one common localization according to HPA. Our goal was to find a specific ratio of the experimental, unknown and predicted evidence type weights that maximizes the number of high confidence interactions in the positive control data set (HPA) and simultaneously maximizes the number of low confidence interactions in the ComPPI data set not containing HPA data. These ensure that the quality of data marked as high confidence will have a good match to the quality of experimentally verified data.

All combinations of the experimental, unknown and predicted evidence type weights were set up from 0 to 1 with 0.1 increments. The kernel density of the interactions were calculated with all these settings (with a bandwidth of 0.01), which gave us the ratio of interactions belonging to a given confidence level compared to the distribution of all the interactions. Finally, the 285 possible kernel density solutions were tested to find the parameter combination that maximizes the number of both the low and high confidence interactions as described above. This resulted in 0.8, 0.7 and 0.3 as the relative evidence type weights for experimental, predicted and unknown data types, respectively (Supplementary Figure S6). Note that this optimization is driven by the reliability of the subcellular localization data, and was not tested using gold standard protein–protein interaction data sets, therefore the Interaction Score reflects the reliability of the interaction in a subcellular localization-dependent but not in an interactome-dependent manner.

**Application examples**

Merging of subcellular localization and interactome data provides several application opportunities: (i) the filtration of localization-based biologically unlikely interactions—where the two interacting proteins have no common localization and (ii) the prediction of possible new localizations and localization-based biological functions (15). Both are important features of ComPPI as illustrated by an example in this section.

*ComPPI-based interaction filtering.* First, we made a systematic search for an example, which highlights the importance of the removal of localization-based biologically unlikely interactions looking for key hubs and bridges, where interaction structure changed the most after the filtering step. Here we calculated the degree distribution of the whole human interactome and the high-confidence interactome (containing 23 265/19 386 proteins and their 385 481/260 829 interactions, respectively) where from the latter biologically unlikely interactions with no common subcellular localizations have already been removed. We also calculated the distribution of the betweenness centrality in the two data sets. After these procedures we manually reviewed the first 20 proteins from the UniProt Swiss-Prot subset (15 258 proteins out of 19 386) with the highest differences in degree and centrality measures (Supplementary Table S5). Enoyl-CoA hydratase (crotonase) had the largest absolute change of degree among the top 20 proteins, thus we selected crotonase as our illustrative example (Figure 4). Crotonase catalyses the second step in the beta-oxidation pathway of fatty acid metabolism (42), and is a key member of the crotonase protein superfamily (43). Beta-oxidation takes place primarily in the mitochondrion (44). Crotonase has only a mitochondrial ComPPI localization with experimental evidence, which is in agreement with its cellular function.

Crotonase has 71 interacting partners in the integrated data set, of which only 8 is present in the mitochondrion, and only 5 have an interaction score equal or higher than 1.8. After the manual review of crotonase neighbours, it turned out that only one of the 8 mitochondrial interactors (mitochondrial Hsp70, (45)) has experimental evidence for mitochondrial localization. Mitochondrial localization of the other 7 interactors is not based on strong evidence, while 63 out of 71 interactors have no known mitochondrial localization at all. Figure 4 shows the interactome of crotonase and its 71 first neighbours containing 428 edges. In the mitochondrial interaction subset only 13 edges remained, while the high-confidence part contains only 10 interactions (Figure 4). Second neighbours of crotonase contain 82% of the interactome, and their network contains 14 803 nodes and 319 305 edges. The filtered mitochondrial network of the second neighbours is much smaller, having only 2107 nodes and 8381 interactions.

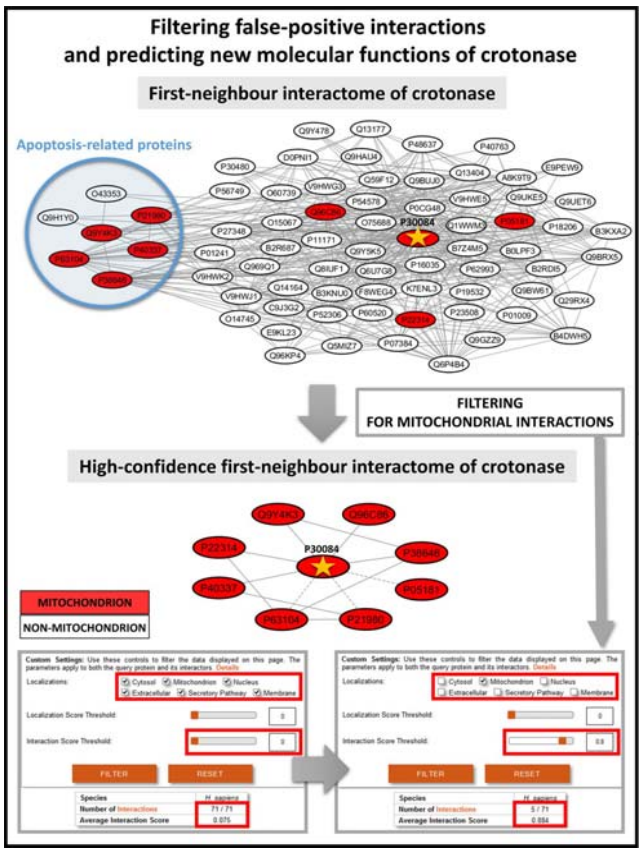

**Figure 4.** Advantages of the ComPPI data set to filter biologically unlikely interactions and to predict compartment-specific, new properties and functions. The figure shows the interactions of crotonase (enoyl-CoA hydratase, UniProt ID: P30084), involved in fatty acid catabolism having a mitochondrial localization, and its first neighbours supported with experimental evidence before and after filtering to mitochondrial localization. Interactions with an Interaction Score below 0.80 are shown with dashed lines. On one hand, out of the original 71 neighbours of crotonase only 8 remain as mitochondrial interacting partners with a significantly higher average Interaction Score than the whole first-neighbour network, which highlights the importance of compartment-specific filtering in the detection of high-confidence interactors in a subcellular localization-dependent manner. On the other hand, the blue circle of the upper left side of the figure shows those cytosolic crotonase interacting partners, which are involved in apoptosis, a recently discovered function of crotonase (47–47). Thus, the very same example also reveals a potential new function of crotonase, which partially involves its unexpected cytosolic localization, which was recently verified experimentally (46).

*ComPPI-based prediction of new or non-conventional functions.* Importantly, 52 out of the 71 interactors, and more specifically, 7 out of the 8 mitochondrial interacting partners of crotonase have cytosolic localization with a localization probability over 0.95. This indicates that crotonase may have a cytosolic localization as well. Indeed, crotonase was shown to be overexpressed and localized in the cytosol in hepatocarcinoma cells, where it contributes to lymphatic metastatis (46). GO (19) biological process term enrichment analysis of the mitochondrial crotonase interacting partners using BiNGO (21) revealed that besides the known function of the crotonase in 'catabolic process' the 'negative regulation of apoptosis' and related terms were also significantly enriched (Supplementary Table S6). In agreement with this, previous studies showed that crotonase is overexpressed in several cancer types (47), and the knockdown of crotonase decreased cell viability and enhanced cisplatin-induced apoptosis in hepatocellular carcinoma (48). The anti-apoptotic effect of crotonase also exists in breast cancer, where its down-regulation potentiates PP2-induced apoptosis (49).

These findings may implicate that the high ratio of 'biologically unlikely' interactions may also be a result of a transient and dynamic cytosolic subcellular localization of crotonase, where the enzyme may be involved in currently not widely crotonase-associated biological processes, such as the inhibition of apoptosis. Importantly, these compartment-specific crotonase functions may be applied as potential therapeutic targets in the treatment of hepatocellular carcinoma or breast cancer.

In summary, the crotonase example shows the utility of ComPPI both (i) to filter low-confidence interactions concentrating on high-confidence subcellular localizations and (ii) to predict unknown biological functions in previously unknown or non-conventional subcellular localizations. Another example of ComPPI-based prediction of potential, novel functions besides crotonase, is Monopolar Spindle 1 protein (MPS1) having a centromere-associated cytosolic localization (50). We identified a number of relatively undiscovered MPS1 functions related to the ComPPI analysis of nuclear MPS1 interactome as detailed in Supplementary Figure S7 and Supplementary Table S6.

## CONCLUSIONS AND FUTURE DIRECTIONS

In summary, ComPPI provides a unique data set for the analysis of protein–protein interaction networks at the subcellular level. The assembly of the integrated ComPPI database with manual curation protocols (Figure 1) provides an improvement of both coverage and data quality. ComPPI subcellular localization data have a novel structure in order to incorporate localizations from different data sources (Figure 3 and Supplementary Figure S4), and to reveal compartment-specific biological functions based on the analysis of the interactomes extended with high-resolution localization data in a hierarchical structure. With the use of the optimized Localization and Interaction Scores (Figure 2) high-confidence interactomes could be created for further investigation in the field of compartment-specific biological processes (15).

Comparison of integrated protein–protein interaction data and the compartmentalized interactome allow the filtering of biologically unlikely interactions, where the interacting partners have no common subcellular localization. Our examples (Figure 4 and Supplementary Figure S7) illustrate that besides filtering, ComPPI has a strong predictive power to find new localizations of the proteins based on the underlying network or to suggest new compartment-specific biological functions. The comprehensive data set for four species gives the opportunity to analyse evolutionary aspects of the compartmentalization, such as the prediction of subcellular localization ortologes ('localogs').

The web interface of ComPPI (http://ComPPI.LinkGroup.hu) provides user-friendly search and download options. Besides the basic Search feature to explore and download the interactions of individual proteins, Advanced Settings could be applied to both query proteins and their interactors. Interactome-wide studies could be applied using the downloadable compartment-specific interactomes or the integrated protein–protein interaction data set, while the integrated subcellular localization data set is also available on the webpage for further analyses.

ComPPI is available at http://ComPPI.LinkGroup.hu, and has an open source code, which allows further improvement and the construction of 'ComPPI-based databases'. ComPPI is a community-annotated resource, which will be continuously enriched by a user community of experts helped by a public issue-tracking system and by feedbacks from the core-team, and will be updated and upgraded annually for minimum 5 years.

We plan to resolve current ComPPI limitations, such as the relatively low amount (29% of total) of experimental subcellular localization entries with the incorporation of newly available experimental data. Future plans include the development of improved gold standard-based Localization and network neighbourhood-based Interaction Scores, as well as further advanced download and search options, such as advanced localization-based network visualization and extended number of output formats.

In summary, the ComPPI-based interactomes introduced here provide a broader coverage, offer highly structured subcellular localization data, as well as offer Localization and Interaction confidence Scores, all in a user-friendly manner. Importantly, ComPPI enables the user to filter biologically unlikely interactions, where the two interacting proteins have no common subcellular localizations, and to predict novel subcellular localization as well as localization-based properties, such as compartment-specific biological functions.

## SUPPLEMENTARY DATA

Supplementary Data are available at NAR Online.

## ACKNOWLEDGEMENT

Authors acknowledge the members of the LINK-Group (http://LinkGroup.hu) and Cellular Network Biology Group (http://NetBiol.elte.hu) for their advice, as well as Judit Gyurkó for the design of the ComPPI website.

## FUNDING

Hungarian Science Foundation [OTKA-K83314]; Fellowship in computational biology at The Genome Analysis Centre (Norwich, UK) in partnership with the Institute of Food Research (University of Norwich, UK); Biotechnological and Biosciences Research Council, UK [to T.K.]; János Bolyai Scholarship of the Hungarian Academy of Sciences [to T.K.]. Funding for open access charge: Hungarian Science Foundation [OTKA-K83314].
*Conflict of interest statement.* None declared.

# Supplementary Material to

## ComPPI: a cellular compartment-specific database for protein-protein interaction network analysis

Daniel V. Veres[1], Dávid M. Gyurkó[1], Benedek Thaler[1,2], Kristóf Z. Szalay[1], Dávid Fazekas[3], Tamás Korcsmáros[3,4,5] and Peter Csermely[1,*]

[1]Department of Medical Chemistry, Semmelweis University, Budapest, Hungary; [2]Faculty of Electrical Engineering and Informatics, Budapest University of Technology and Economics, Budapest, Hungary; [3]Department of Genetics, Eötvös Loránd University, Budapest, Hungary; [4]TGAC, The Genome Analysis Centre, Norwich, UK; [5]Gut Health and Food Safety Programme, Institute of Food Research, Norwich, UK;
*To whom correspondence should be addressed. Tel: +361-459-1500; Fax: +361-266-3802; Email: csermely.peter@med.semmelweis-univ.hu; The authors wish it to be known that, in their opinion, the first 2 authors should be regarded as joint First Authors.

**Table of Contents**

Supplementary Figure S1. **The role of multiple subcellular localizations on the example of IGFBP-2 and HIF-1 Alpha proteins.**

Panel **(A)** depicts the compartment-specific functions of IGFBP-2. IGFBP-2 is a secreted (dark blue), predominantly extracellular (green) protein involved in insulin growth factor (IGF) signalling (1). IGFBP-2 is also localized in the nucleus (orange), where it is required for the activation of VEGF-mediated angiogenesis (2). This binary function of IGFBP-2 turned out to be an important prognostic biomarker in several cancer types, such as breast cancer (3). Panel **(B)** shows another important example, HIF-1 Alpha, which is a member of the hypoxia inducible factor (HIF) family, and has a key importance in the maintenance of cellular oxygen homeostasis, particularly in the response to hypoxia (4). In order to act as a transcription factor, HIF-1 Alpha has to translocate from the cytosol (light blue) into the nucleus (orange). HIF-1 Alpha and the aryl hydrocarbon receptor nuclear translocator (ARNT) form a heterodimer in the nucleus, which is required for their stable nuclear association (5). This regulated transport has key role in HIF-1 Alpha activity control both in healthy conditions (6) and in the development of cancer (7).

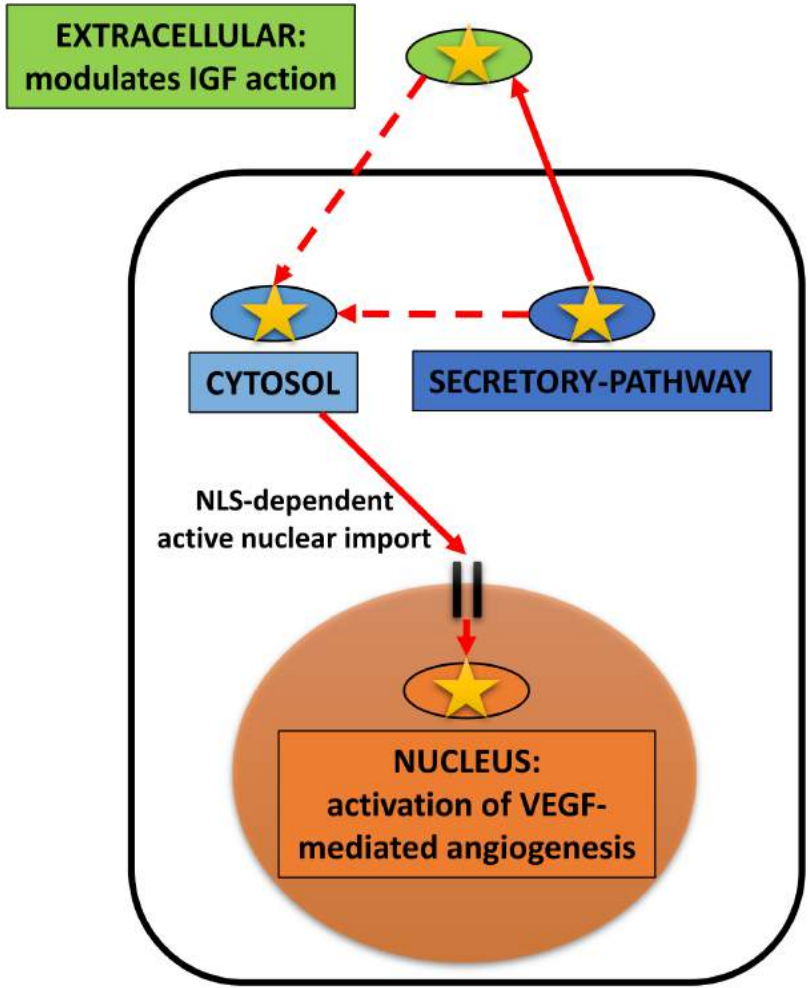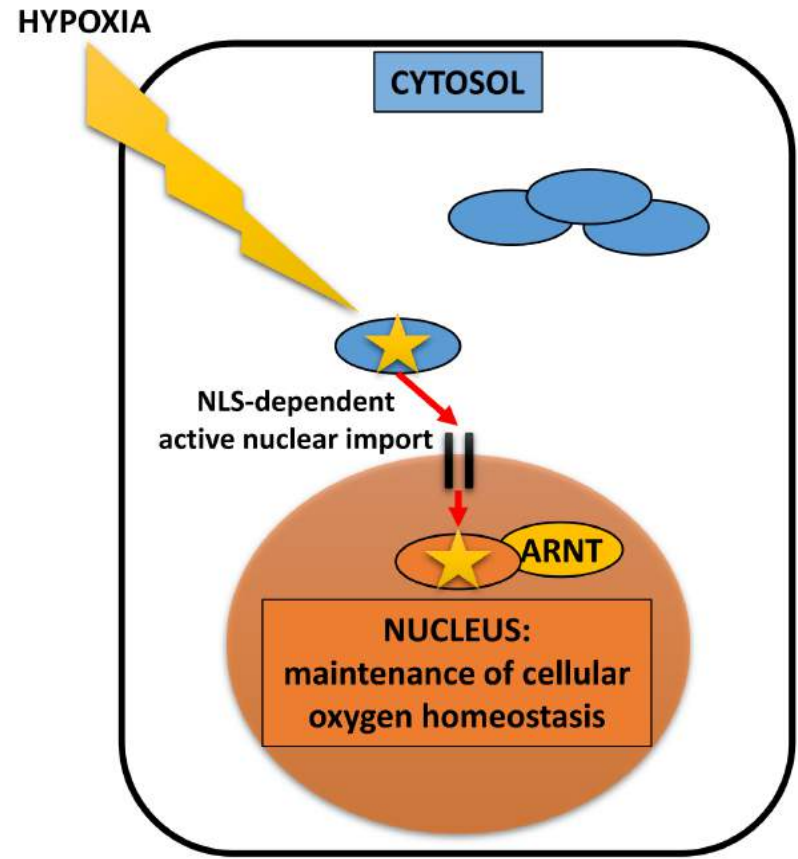

Supplementary Figure S2. **The localization tree used for the ComPPI subcellular localization data integration.**

ComPPI subcellular localization data are assembled from several sources with different naming conventions and resolution, therefore a standardization step is required before integration. We built a non-redundant, hierarchical subcellular localization tree based on more than 1,600 Gene Ontology (8) cellular component terms manually. The tree structure is needed to separate the high resolution subcellular localization data unequivocally to the 6 major cellular compartments of low resolution for further analysis (Supplementary Table S4). On the example shown we demonstrate the tree structure on a segment of the hierarchical subcellular localization tree with the nucleus in focus. The high resolution data, such as the components of the nuclear pore complex, are highlighted in orange at the bottom part of the figure. The whole subcellular localization tree is downloadable (http://www.linkgroup.hu/pic/loctree.svg), or available online:
http://bificomp2.sote.hu:22422/comppi/files/c6f5587545cca11928563be40a0f8d6bf4bf45b2/databases/loctree/loctree.textile

# The structure of the localization tree

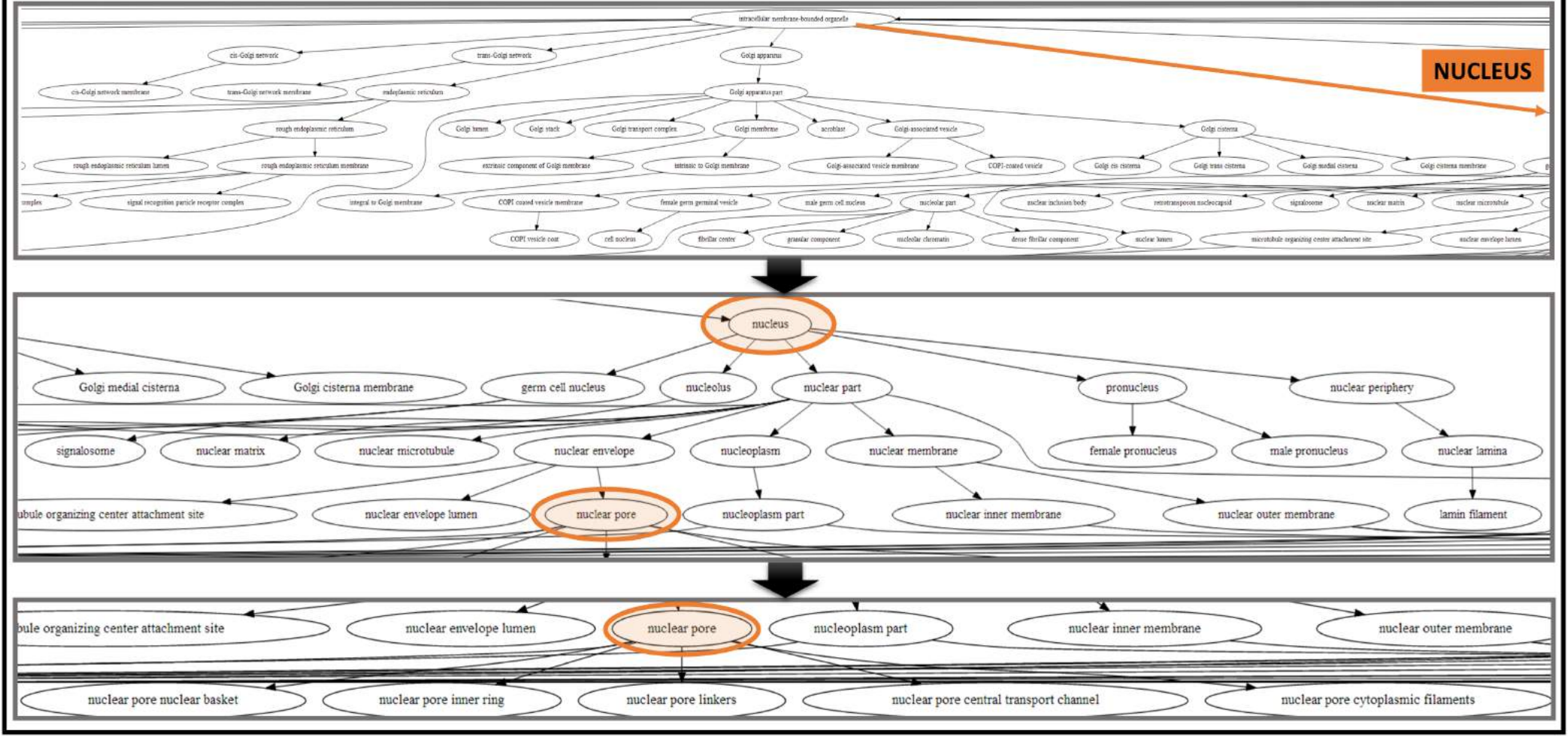

Supplementary Figure S3. **The overlap between subcellular localization and protein-protein interaction input databases used in the assembly of ComPPI.**

Panels **(A)** and **(B)** show the overlap between the source subcellular localization and protein-protein interaction (interactome) databases graphically for all the data and after filtering by species as well. Nodes in the networks symbolize the source databases, their size is proportional to the number of proteins originated from the given database. Edges connecting the nodes represent the overlap, their width correlates with the number of common proteins in the interconnected databases. Interestingly, these graphs are fully connected, e.g. all databases share proteins with each other (except DroID, a *D. melanogaster*-specific database, which has no overlap with databases lacking fruit fly data). The reason is usually the redundancy between the databases (for example MINT and IntAct share their data as part of the MIntAct project (9)).

Panels **(C)** and **(D)** display numerically the same data as Panel **(A)** and **(B)**, respectively: the number of proteins (Y axis) are plotted for every input database (X axis), each bar represents an edge from Panel **(A)** or **(B)**. Significant differences can be observed in the amount of source data provided by the input databases.

Panel **(E)** shows the number of proteins (Y axis) per source database (X axis) for the given species compared to the number of proteins in ComPPI (highlighted in orange). The number of common proteins in all source databases, in databases with more than 5000 proteins, and only in interactome databases are also represented on the charts. Panel **(F)** shows the number of interactions (Y axis) for the given species per interactome source databases (X axis) compared to the number of proteins in ComPPI (highlighted in orange).

Abbreviations: HPA, Human Protein Atlas; H.Proteinpedia, Human Proteinpedia

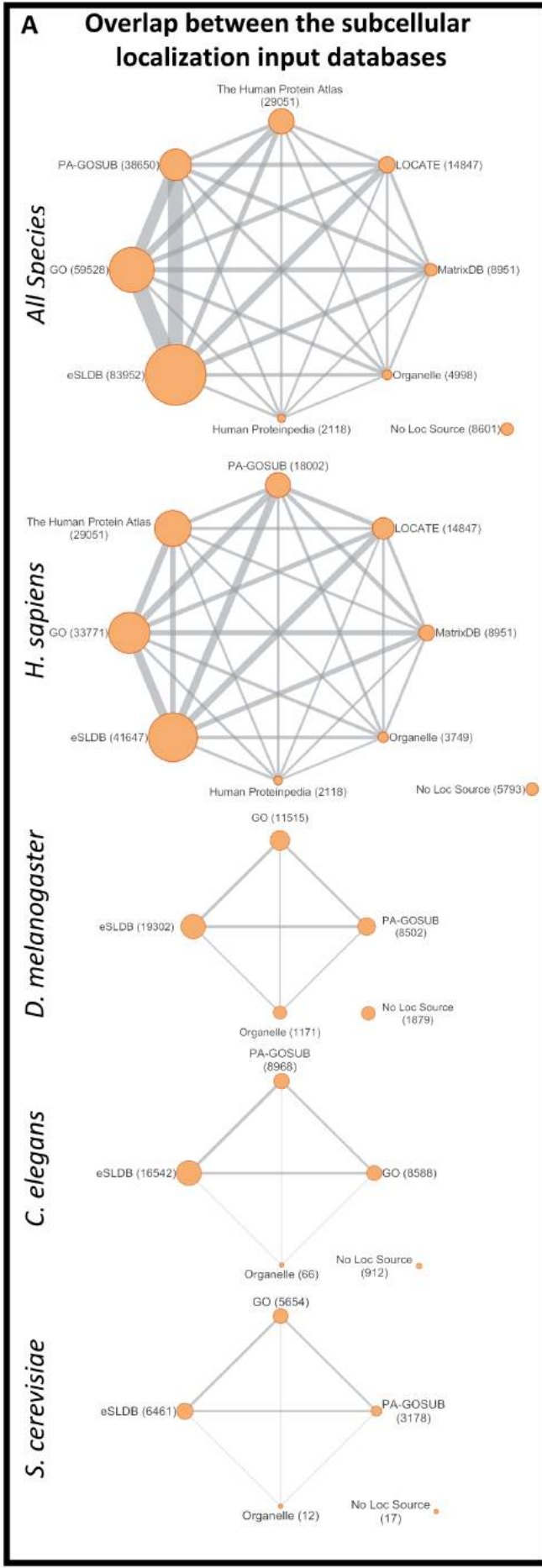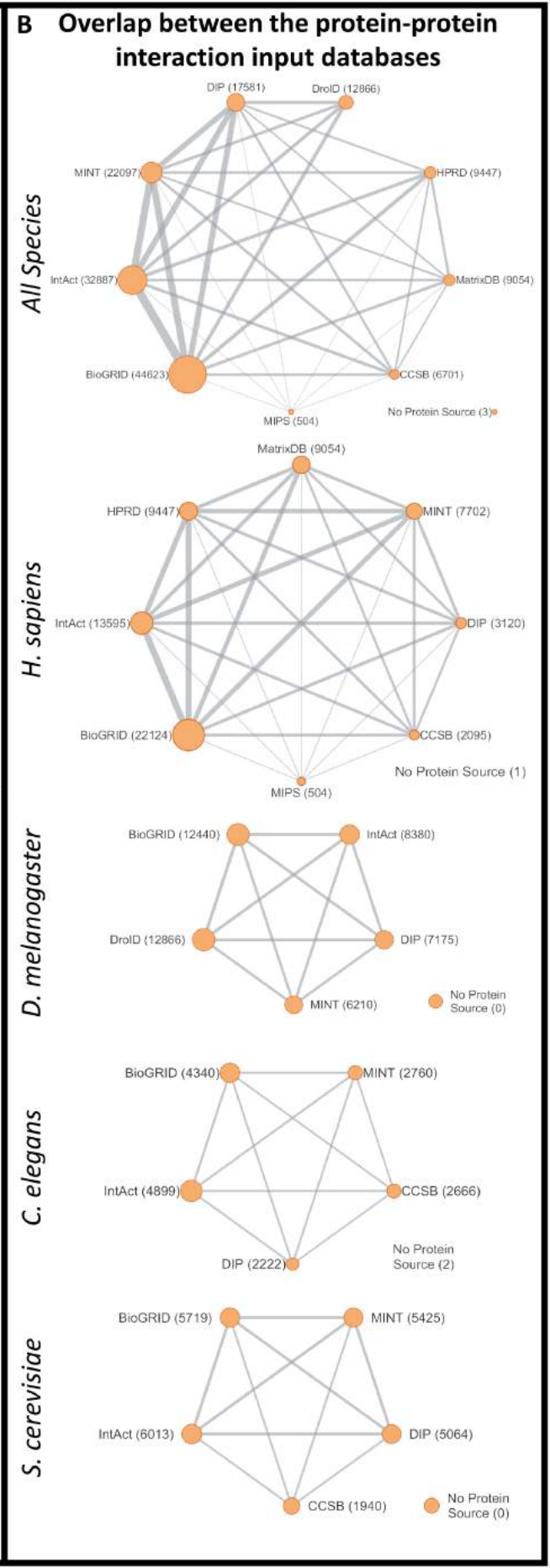

## C. Overlap between the subcellular localization input databases - data

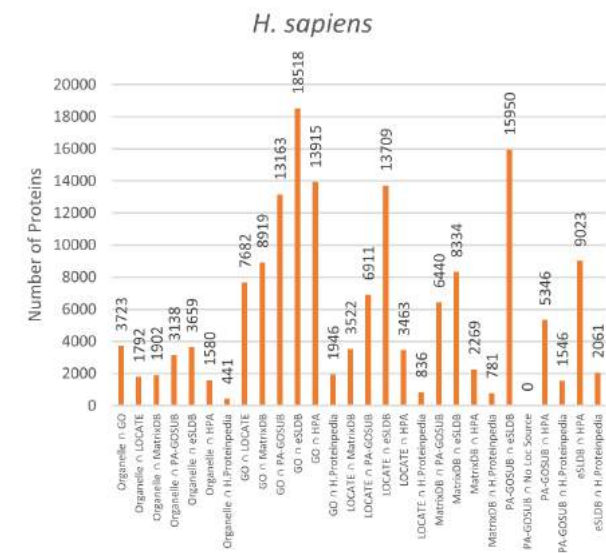
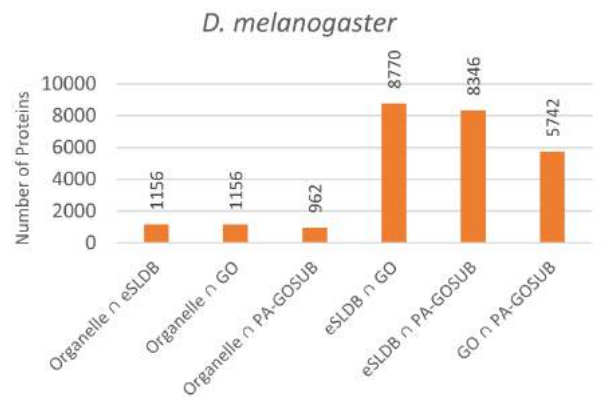
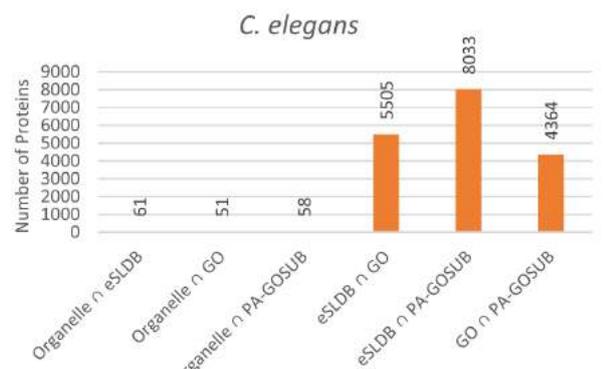
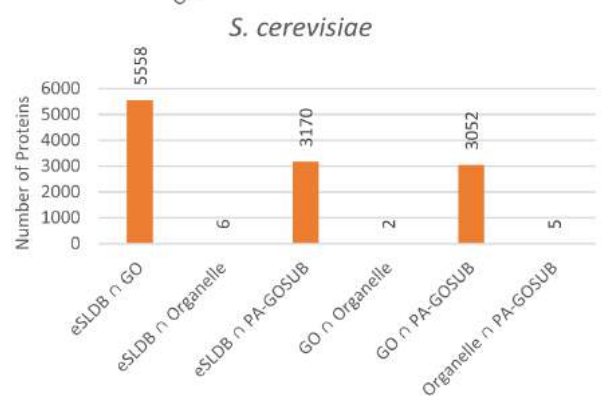

## D. Overlap between the protein-protein interaction input databases - data

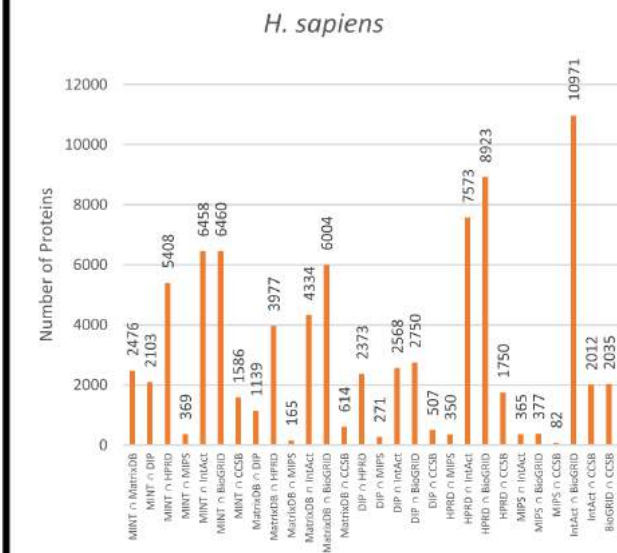
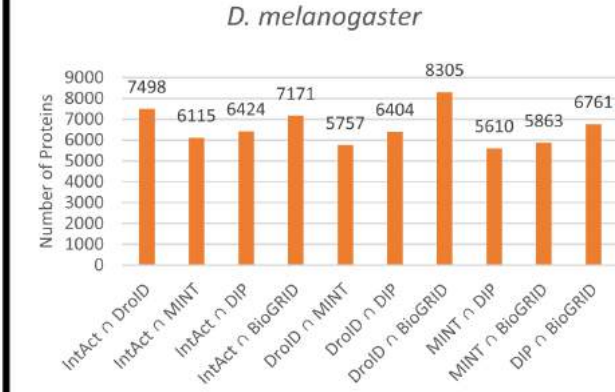
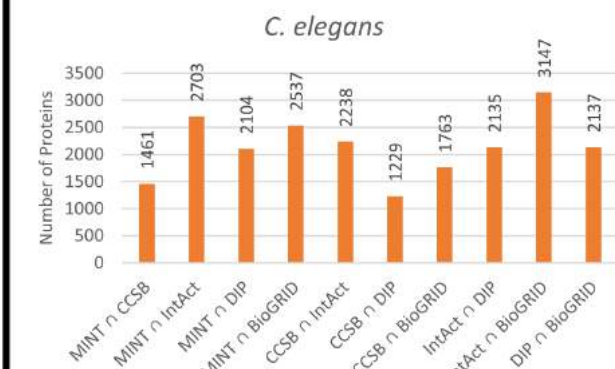
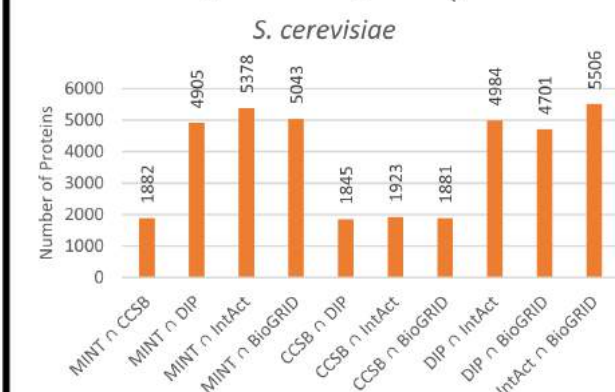

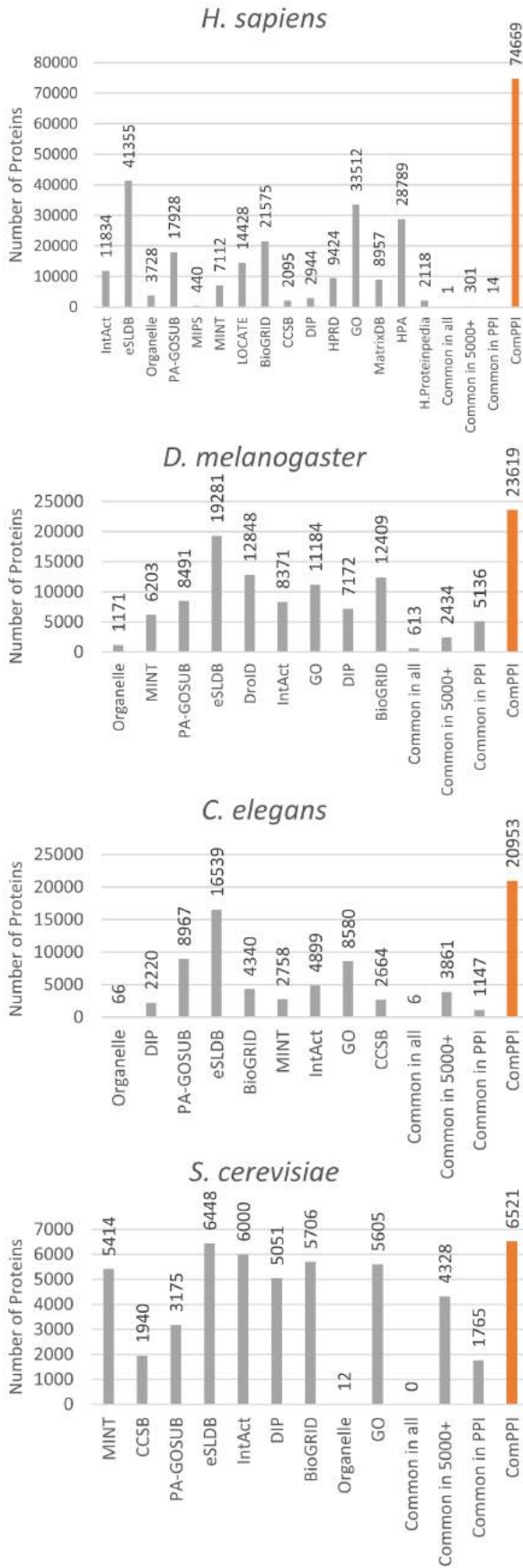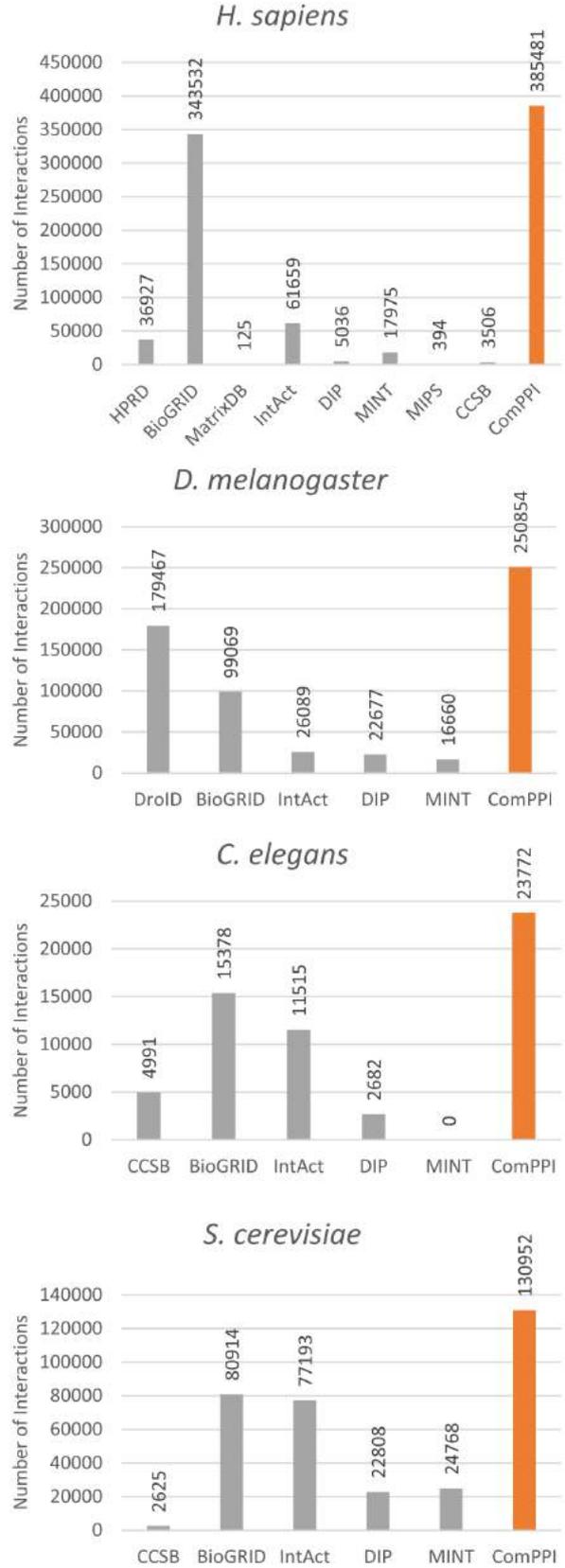

Supplementary Figure S4. **Advantages of multiple localization evidence.**

The Venn-diagrams represents the number of proteins with experimental, unknown, predicted, or mixed localization evidence in the integrated ComPPI subcellular localization dataset and in the sub-dataset of ComPPI containing GO-based localization data for its proteins. Interestingly, the number of proteins with only unknown localization evidence is much lower in ComPPI compared to its GO-subset (57%), while the number of proteins with experimental (264%) and mixed evidence (376%) is higher, than the estimation based on the number of proteins in ComPPI compared to the GO sub-dataset (203%). These findings implicate that the integration of subcellular localization data from different sources having different localization evidence types increases not only the quantity of the data, but their reliability as well.

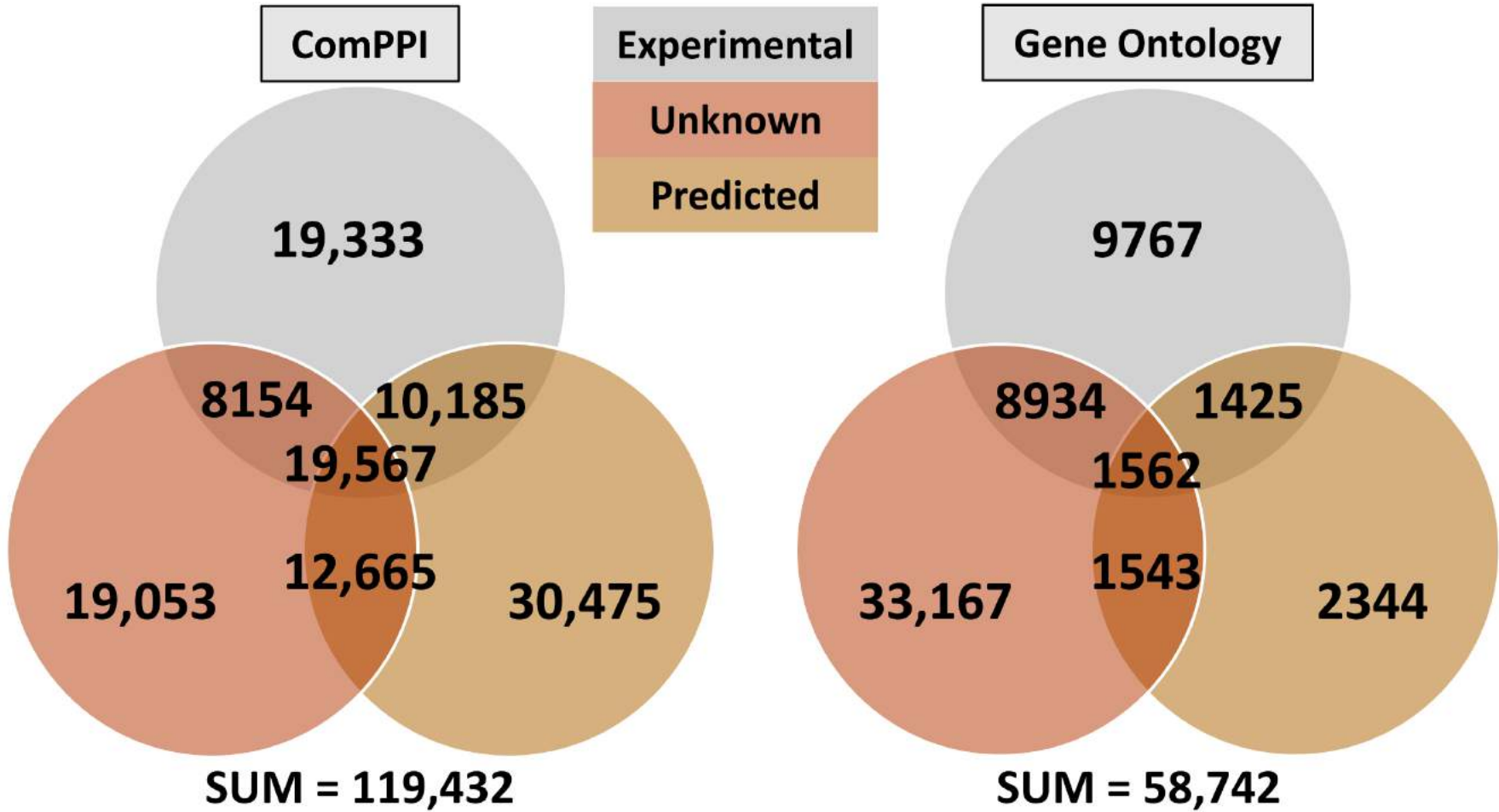

Supplementary Figure S5. **The distribution of the ComPPI Localization and Interaction Scores.**

Panel **(A)** shows the distribution histogram of the Localization Scores for the 4 species of ComPPI separated by the 6 major cellular compartments containing all subcellular localization data. The X axis represents the confidence intervals of the Localization Score between 0 and 1 by 0.1 increments, while the Y axis shows the number of proteins belonging to the given compartment. On panel **(B)** the distribution histogram of the Interaction Scores is shown for the 4 species of ComPPI containing the integrated protein-protein interaction data. The X axis depicts the confidence intervals of the Interaction Score between 0 and 1 by 0.1 increments, while the Y axis shows the number of interactions in the given confidence interval. See more about the Localization and Interaction Scores in Figure 2 and Supplementary Figure S6.

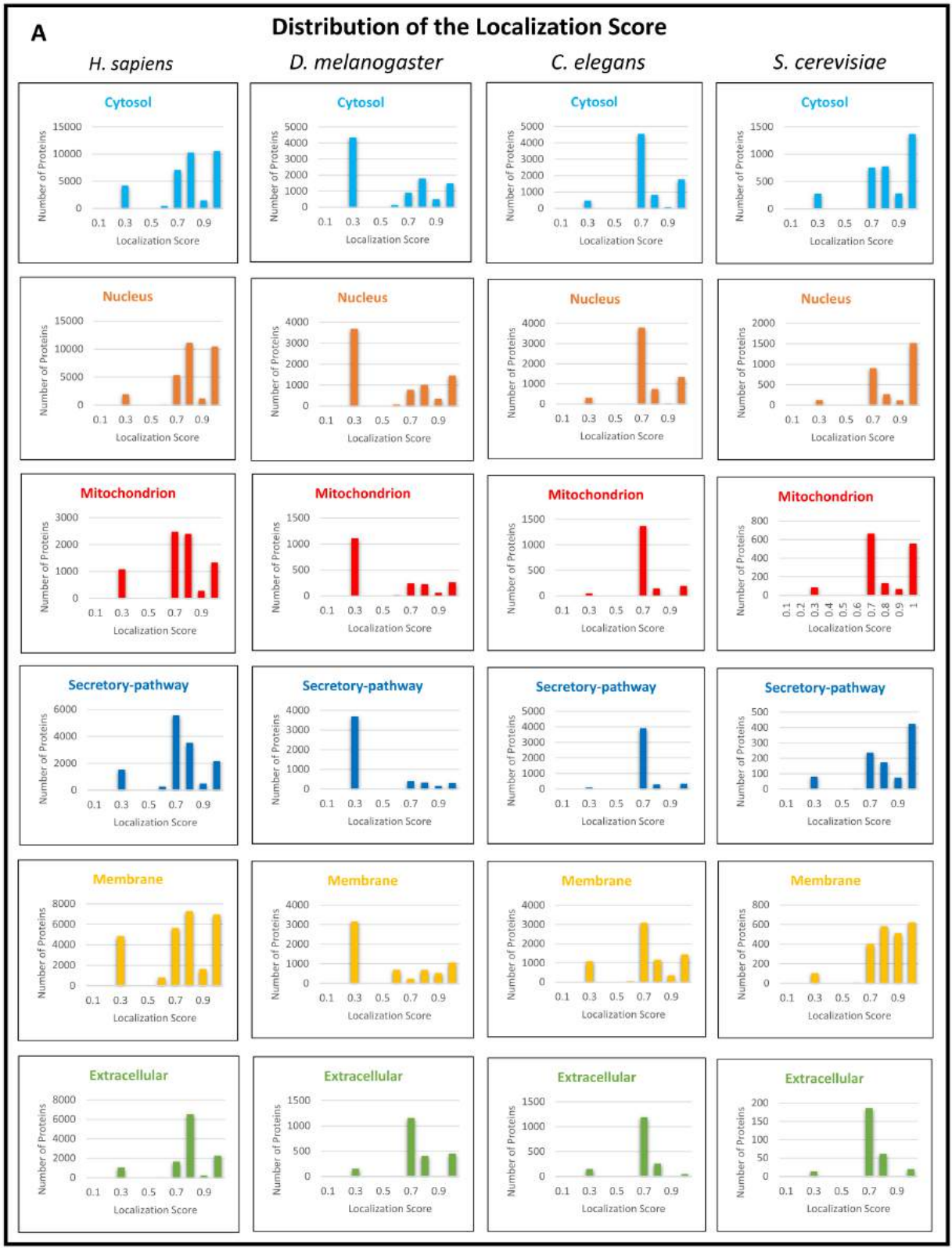

**B**                Distribution of the Interaction Score

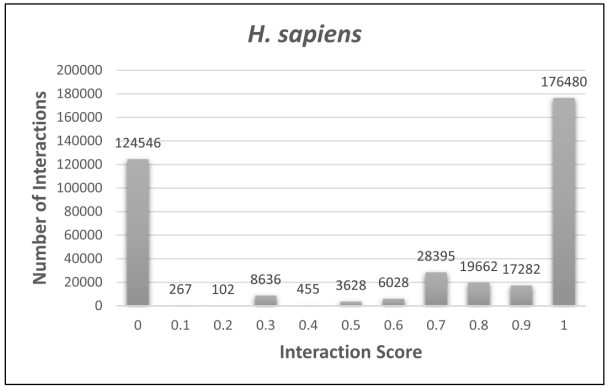
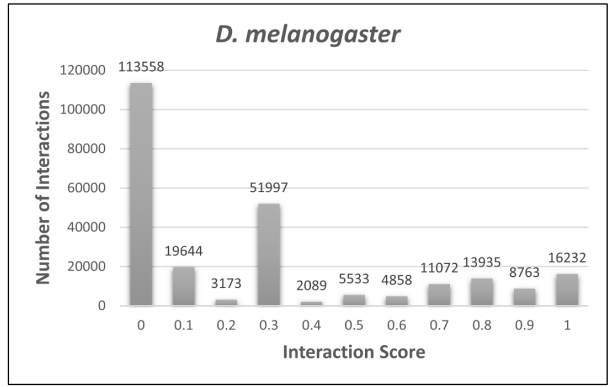
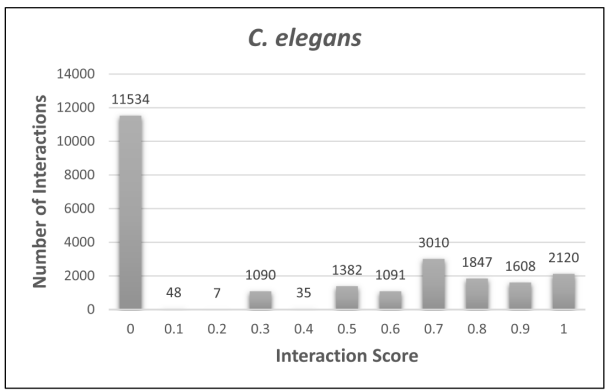
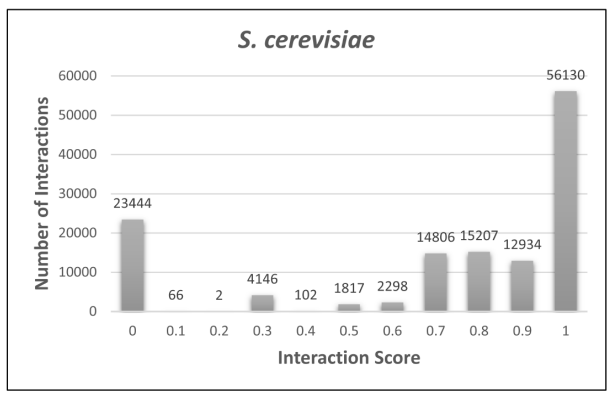

Supplementary Figure S6. **Optimization of the ComPPI localization evidence type weights.**

The flowchart on panel **(A)** shows the steps of ComPPI localization evidence type weight optimization. Our goal was to define the ratio of the localization evidence type weights to each other, and to set up a threshold, above which the interactions are considered trustworthy, i.e. as 'high confidence' as the known experimental data. Localization evidence types can be experimental, unknown and predicted. For more details, see the "Localization and Interaction Scores" section in the main text. Each evidence type has a parameter that defines its weight against the other two types, termed as localization evidence type weights. To define these weights, we built a positive-control dataset containing only those interactions, where both interactors are localized in the same subcellular compartment, and this information was supported with experimental evidence from the Human Protein Atlas database (HPA, 32). We compared the positive-control dataset (56,160 interactions, in green) to all the interactions from the ComPPI dataset excluding the subcellular localization data from HPA (790,269 interactions, in grey). Based on the fact that experimentally validated entries are the most reliable, while localization entries coming from unknown or predicted origin are less reliable, we set the following order of evidence type weights: experimental > predicted AND experimental > unknown as the two requirements of the optimization process. All combinations of the experimental, unknown and predicted localization evidence type weights were set up from 0 to 1 with 0.1 increments. The kernel density of the interactions were calculated using all these settings, which gave us the ratio of interactions belonging to a given confidence level compared to the distribution of all the interactions. Our goal was to find a specific ratio of the experimental, unknown and predicted localization evidence type weight parameters that maximizes the number of high confidence (HQ) interactions in the positive control dataset (HPA) and simultaneously maximizes the number of low confidence (LQ) interactions in the ComPPI dataset not containing HPA data. These ensure that the quality of data marked as high confidence will match the quality of experimentally verified data. We calculated the 95% confidence threshold (in red) for the positive-control dataset in order to separate our data into high- and low-confidence subsets for all the 285 solutions. The 285 possible kernel density solutions were tested to find the parameter combination that maximizes the number of both the low and high confidence interactions as described above. The number of HQ HPA interactions is much lower than the number of LQ interactions, therefore these had to be normalized to avoid any statistical bias. The normalized values were multiplied to obtain a single value representing the combination of the specific experimental, unknown and predicted localization evidence type weights. We simultaneously maximized the number of interactions in the high-confidence positive control dataset (HQ HPA) and the number of interactions in the low-confidence subset of all the ComPPI interactions not containing HPA (LQ ALL):

$$\frac{NumberOfInteractions \in HQ\ HPA}{MaximalNumberOfInteractions \in HQ\ HPA} * \frac{NumberOfInteractions \in LQ\ ALL}{MaximalNumberOfInteractions \in LQ\ ALL}$$

(Eq.1)

The single values were ranked and the largest one was selected, that represented the localization evidence type weights of 0.8, 0.7 and 0.3 for the experimental, predicted and unknown localization evidence types, respectively. Panel **(B)** illustrates the distribution of the Interaction Score after optimization. The X axis shows the Interaction Score, while the Y axis represents the kernel density of the interaction distribution. The figure shows that Interaction Score distribution for the positive-control dataset (56,160 interactions, in green) and for the interactions not containing HPA localizations (790,269 interactions, in grey), which had the largest value of Eq. 1, i.e. which had the optimal set of localization evidence type weights out of the possible 285 representations. The number of high-confidence interactions in the positive-control dataset with a confidence threshold at 95% equalled 48,567, while the number of high-confidence interactions in the ComPPI dataset excluding HPA was 241,076 (30% of total). The optimal distribution shown represented the 0.8, 0.7 and 0.3 localization evidence type weight set for the experimental, predicted and unknown localization evidence types, respectively. The relatively low weight (0.8) of the experimental evidence type means that (i) a single evidence does not result in highly trusted subcellular localization and (ii) at least two pieces of experimental evidence are required to have a localization score above 95%. The relatively high weight (0.7) of the predicted evidence type is in agreement with the high reliability of subcellular localization prediction methods, while the low weight (0.3) of the unknown evidence type highlights the need of the validation of data origin. These facts taken together highlight the importance of data integration and allow a strong filtering of the ComPPI dataset resulting in reliable high-confidence interactions.

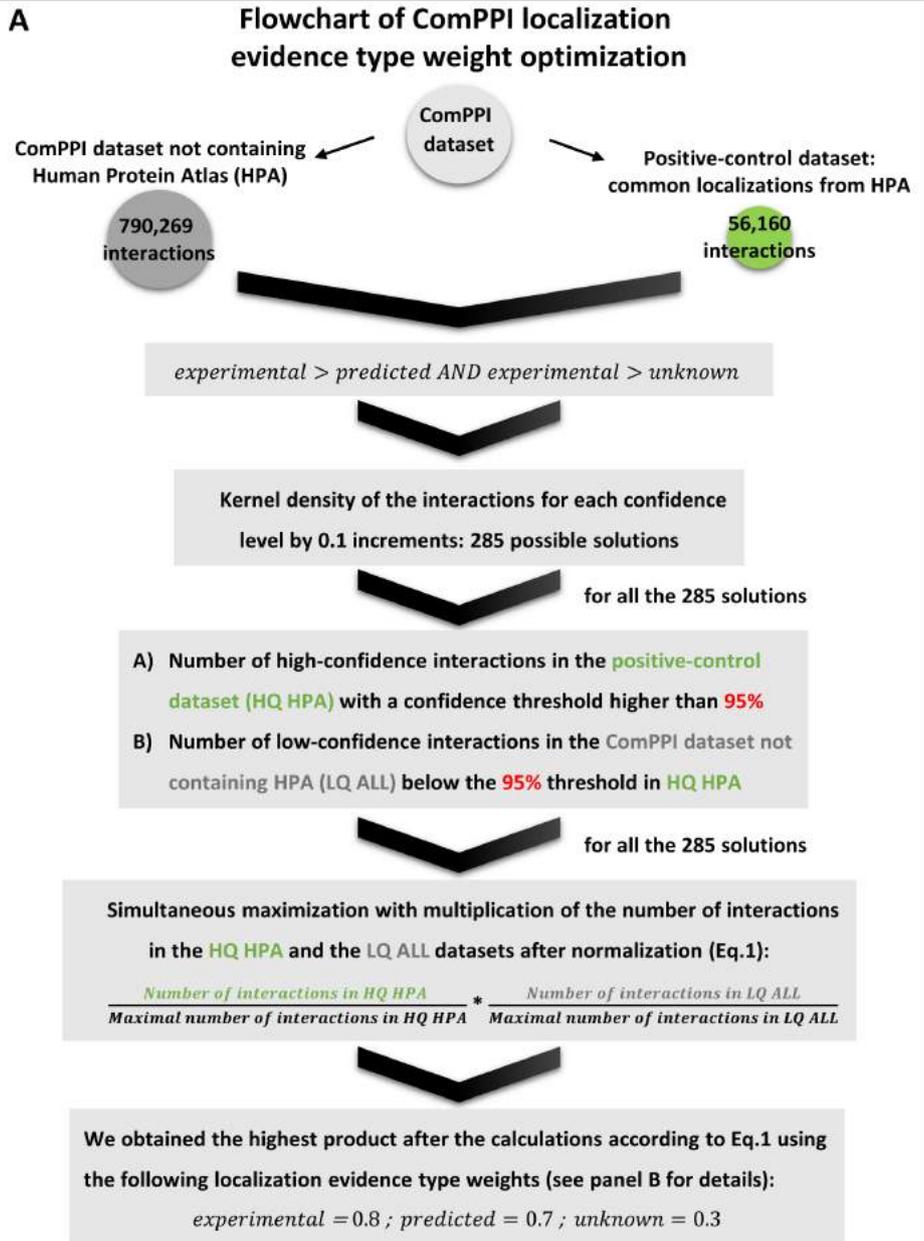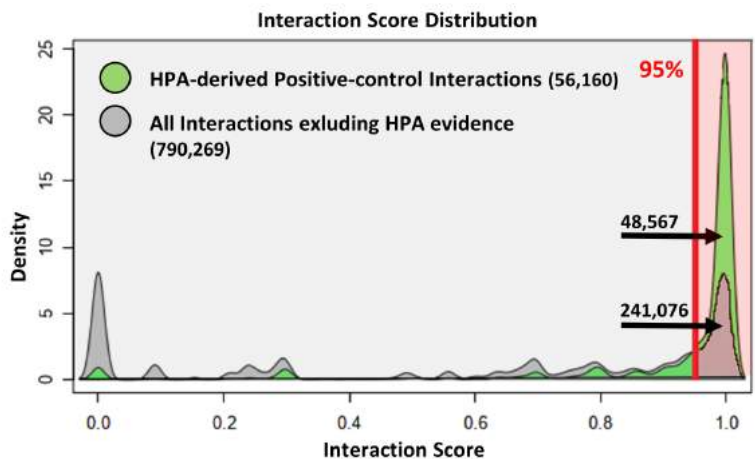

Supplementary Figure S7. **Advantages of ComPPI: prediction of compartment-specific, new properties and functions of MPS1.**

The figure shows another example of ComPPI-based prediction of potential, novel functions besides the example of crotonase detailed in the main text. We choose a well-known dual-specific protein kinase, the Monopolar Spindle 1 protein (MPS1, UniProt ID: P33981) having a centromere-associated cytosolic localization (10) and relatively undiscovered functions in its nuclear subcellular localization resulting in incomplete knowledge of the effect of its inhibitors. MPS1 is a member of the spindle assembly checkpoint complex (11), and its inhibition causes aneuploidy *via* mitotic arrest resulting in apoptosis (12). Importantly, MPS1 also has a nuclear localization (13), which is independent from its inhibition (14). MPS1 inhibitors are potent anticancer drugs with well-characterized pharmacological effects (15). These findings raise the idea that MPS1 may have still unknown effects in the nucleus after its inhibition. However, the exact role of MPS1 in the nucleus remains rather unclear. As one of the sporadic pieces of related evidence MPS1 mediates epigenetic functions, such as the regulation of the chromatin organization through the phosphorylation of Condensin-2 (16). Additionally, MPS1 has a still unknown function attached to the nucleoplasmic side of the nuclear pore complex (13). In ComPPI has MPS1 has 40 interacting partners. Supplementary Figure S7 shows the interactions of the MPS1 kinase and its first neighbours before and after filtering to nuclear localization using an Interaction Score threshold of 0.90. 28 of the total 40 of its interactors (70%) have nuclear localization. This result proposes a more extensive nuclear function of MPS1 than previously suspected. To assess the putative nuclear functions of MPS1, we built the interactome of the first and second neighbours of MPS1, and filtered them for nuclear interactions. Gene Ontology (8) enrichment analysis using BiNGO (17) showed that besides the already known MPS1-related biological functions, such as 'mitotic cell cycle checkpoint' or 'chromosome separation', additional functions, such as 'cellular component organization', 'organelle organization' and related terms also showed a significant enrichment (Supplementary Table S6). This is in agreement with the earlier suggestion that MPS1 may have a role in nuclear assembly and cellular component reorganization during mitosis possibly related to its localization in the nuclear pore complex (13). The assessment of the function of first and second MPS1 neighbours revealed 3 second neighbours playing key role in nuclear assembly, (Lamin-B2 (18), LAP2 (19) and Emerin (20)), and 3 others involved in epigenetic regulation of chromatin condensation and decondensation (Aurora kinase B (21) and C (22), as well as histone acetyltransferase p300 (23)). Lamin B2 is connected to MPS1 through its 2 first neighbours, where one of them is the VCP protein (highlighted in green circle). VCP is involved in DNA-damage response (24), and is known as a regulator of the nuclear envelope reassembly (25). Additionally, three of the first neighbours of MPS1 are members of the anaphase promoting complex/cyclosome (APC/C) (highlighted in red circle), a complex having key role in the maintenance of the spindle assembly checkpoint by the ubiquitination of the Cdc20 (26). Furthermore, APC/C mediates the degradation of MPS1 and helps the coordination of the re-entry to the cell cycle during environmental stress (27), which can be a rescue mechanism for cancer cells as well. The potential function of MPS1 in nuclear reassembly needs further investigation and experimental validation. Despite this uncertainty, this example also highlights the importance of the analysis of compartment-specific sub-networks and the predictive power of ComPPI.

# Prediction of new nuclear functions of the kinetochore/cytoplasmic MPS1 kinase

## First-neighbour interactome of MPS1

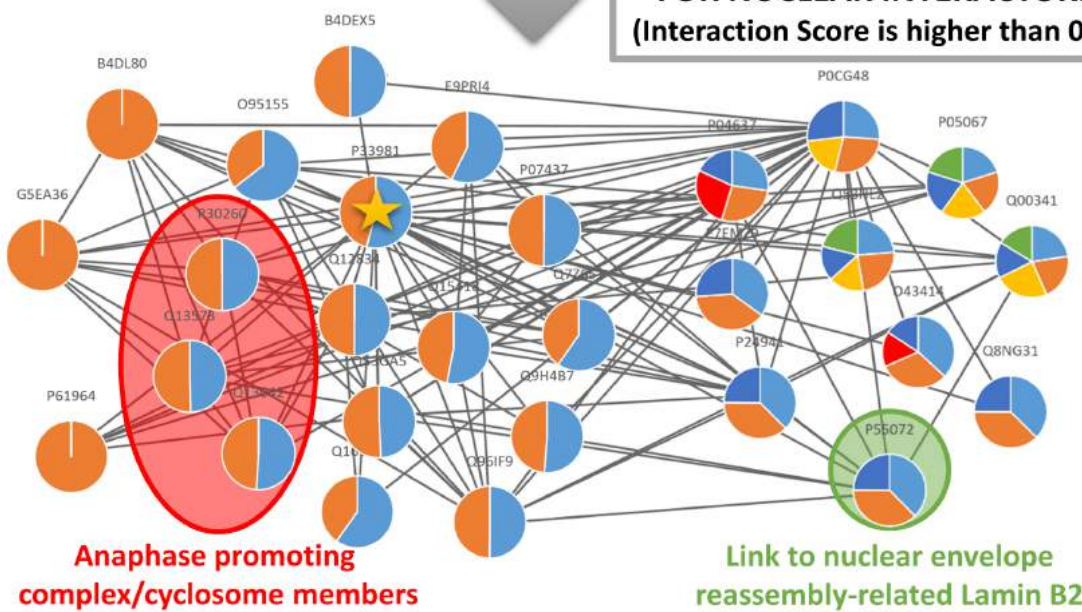

**FILTERING FOR NUCLEAR INTERACTORS** (Interaction Score is higher than 0.9)

**Anaphase promoting complex/cyclosome members**

**Link to nuclear envelope reassembly-related Lamin B2**

Supplementary Table S1. **Comparison of ComPPI with similar tools and databases.**

ComPPI is a unique resource, with an integrated protein-protein interaction dataset enriched by the subcellular localization of the interactors, in order to perform more reliable compartment-specific analysis of biological processes. We present here only those key examples from the wide set of tools, databases and scoring algorithms, that had similar purposes to those of ComPPI. Besides a brief summary some key advantages and disadvantages of the resources, and references to them are listed. These tools and databases are open source, and mainly use Gene Ontology (8) based subcellular localization annotations. See Figure 3 in the main text and Supplementary Figures S3 and S4 for the advantages of the ComPPI structure and the comparison to the integrated source databases.

|  | **NAME** | **SUMMARY** | **ADVANTAGES** | **DISADVANTAGES** | **PUBMED ID** |
|---|---|---|---|---|---|
| **TOOLS** | Cerebral | Layout of molecular interaction networks using subcellular localization annotation. | - easy-to-use Cytoscape plugin<br>- simple visualization of the subcellular organization<br>- input data could be not only GO annotation | - input data are not defined<br>- difficult to visualize proteins with multiple localizations<br>- export option is only for the network image | 17309895 |
|  | BiNGO | Enrichment analysis of GO terms in molecular networks with user-friendly visualization options. | - Cytoscape plugin<br>- enrichment analysis and visualization of GO cellular component terms<br>- available for several species | - only GO annotations<br>- input network has to be imported by the user<br>- no export options for the results | 15972284 |
|  | BioNetBuilder | Interface to create biological networks integrated from several databases. | - Cytoscape plugin<br>- source databases connect via interfaces<br>- the user could set database parameters before integration<br>- ~ 1000 different species | - limited number of input databases<br>- pure-integration without manual follow-up | 17138585 |
|  | CmPI | Reconstruction of the subcellular organization of the proteins in a 3D virtual cell. | - visualization of the interactome based on subcellular localization in 3D<br>- wide options for input data<br>- downloadable software | - filtering for subcellular localization is not available, only the visualization | 23427987 |
| **DATABASES** | HitPredict | Integrated protein-protein interaction dataset with predicted confidence levels. | - interactomes for 9 species with confidence score<br>- confidence score based on structural data, GO annotation and homology<br>- download option of high-confidence interactomes | - only GO-based localization information<br>- compartment specific networks are not available<br>- only 3 input databases (IntAct, BioGRID and HPRD) | 20947562 |
|  | InterMitoBase | Integrated high quality interactome of the human mitochondrion. | - high-confidence interactome for mitochondrial proteins<br>- integration of KEGG, HPRD, BioGRID, DIP and IntAct<br>- graph visualization | - only mitochondrial network<br>- only 490 mitochondrial proteins | 21718467 |
|  | MatrixDB | Manually curated high-quality interaction database for extracellular proteins and other molecules. | - high-quality interactome for the extracellular matrix<br>- subcellular localization data for membrane, secreted and extracellular proteins<br>- several download options | - only for the extracellular matrix<br>- confidence score for the interactions is not available | 20852260 |
| **SCORING ALGORITHM** | PRINCESS | Online interface for human protein-protein interaction confidence evaluation. | - complex scoring system available online<br>- network topology is also included during the analysis | - only for human<br>- subcellular co-localization is based only on GO annotation | 18230642 |

Supplementary Table S2. **Statistics of the ComPPI database.**

The table shows the brief statistics of the ComPPI dataset after the integration of different sources. ComPPI contains three types of downloadable datasets, (i) the compartmentalized interactome, where the interacting proteins have at least one common subcellular localization, (ii) the integrated protein-protein interaction dataset without localization information, and (iii) the integrated subcellular localization dataset. 'Summary Statistics' shows the summary of the dataset for all 4 species. The detailed statistics of the three dataset are available for each species for all localizations together and per each major cellular component. Only the average Localization and Interaction Scores are represented in this table. For more details about the distribution of Localization and Interaction Scores see Supplementary Figure S5.

| SPECIES | DATA TYPE | NUMBER OF PROTEINS | NUMBER OF MAJOR LOCALIZATIONS | AVERAGE LOCALIZATION SCORE | NUMBER OF INTERACTIONS | AVERAGE INTERACTION SCORE |
|---|---|---|---|---|---|---|
| | **COMPARTMENTALIZED INTERACTOME** | | | | | |
| | *Summary Statistics* | 42829 | 86874 | 0.76 | 517461 | 0.76 |
| | **INTEGRATED PROTEIN-PROTEIN INTERACTION DATASET** | | | | | |
| | *Summary Statistics* | 53168 | - | - | 791059 | 0.49 |
| | **INTEGRATED SUBCELLULAR LOCALIZATION DATASET** | | | | | |
| ***All Species*** | *Summary Statistics* | 119432 | 195815 | 0.73 | - | - |
| | **COMPARTMENTALIZED INTERACTOME** | | | | | |
| | *All Localizations* | 19386 | 47761 | 0.82 | 260829 | 0.88 |
| | *Cytosol* | 12801 | 35498 | 0.82 | 185012 | 0.91 |
| | *Mitochondrion* | 1937 | 6202 | 0.81 | 9433 | 0.92 |
| | *Nucleus* | 10820 | 27540 | 0.83 | 156601 | 0.93 |
| | *Extracellular* | 5848 | 20983 | 0.83 | 29725 | 0.96 |
| | *Secretory Pathway* | 5114 | 18299 | 0.81 | 27425 | 0.93 |
| ***H. sapiens*** | *Membrane* | 8408 | 27800 | 0.82 | 57509 | 0.91 |
| | *All Localizations* | 13332 | 20970 | 0.59 | 137011 | 0.46 |
| | *Cytosol* | 7037 | 12715 | 0.62 | 81199 | 0.50 |
| | *Mitochondrion* | 911 | 1803 | 0.61 | 3717 | 0.53 |
| | *Nucleus* | 5507 | 9239 | 0.61 | 48279 | 0.51 |
| | *Extracellular* | 737 | 1764 | 0.61 | 1482 | 0.69 |
| | *Secretory Pathway* | 2276 | 4726 | 0.58 | 10541 | 0.44 |
| ***D. melanogaster*** | *Membrane* | 2955 | 5742 | 0.64 | 15758 | 0.49 |
| | *All Localizations* | 4221 | 7369 | 0.77 | 12233 | 0.68 |
| | *Cytosol* | 2369 | 4664 | 0.77 | 6039 | 0.73 |
| | *Mitochondrion* | 181 | 416 | 0.75 | 156 | 0.71 |
| | *Nucleus* | 1995 | 3685 | 0.77 | 5849 | 0.71 |
| | *Extracellular* | 68 | 151 | 0.73 | 80 | 0.47 |
| | *Secretory Pathway* | 809 | 1752 | 0.75 | 1189 | 0.70 |
| ***C. elegans*** | *Membrane* | 629 | 1295 | 0.79 | 1269 | 0.70 |
| | *All Localizations* | 5890 | 10774 | 0.82 | 107387 | 0.84 |
| | *Cytosol* | 3374 | 7106 | 0.81 | 69698 | 0.85 |
| | *Mitochondrion* | 1407 | 2969 | 0.78 | 10668 | 0.81 |
| | *Nucleus* | 2819 | 5554 | 0.81 | 51035 | 0.89 |
| | *Extracellular* | 147 | 388 | 0.78 | 230 | 0.83 |
| | *Secretory Pathway* | 891 | 2264 | 0.82 | 4560 | 0.91 |
| ***S. cerevisiae*** | *Membrane* | 1876 | 4077 | 0.83 | 11891 | 0.87 |

| INTEGRATED PROTEIN-PROTEIN INTERACTION DATASET | | | | | | |
|---|---|---|---|---|---|---|
| *H. sapiens* | - | 23266 | - | - | 385481 | 0.60 |
| *D. melanogaster* | - | 17379 | - | - | 250854 | 0.25 |
| *C. elegans* | - | 6298 | - | - | 23772 | 0.35 |
| *S. cerevisiae* | - | 6228 | - | - | 130952 | 0.69 |
| INTEGRATED SUBCELLULAR LOCALIZATION DATASET | | | | | | |
| *H. sapiens* | *All Localizations* | 71271 | 123225 | 0.76 | - | - |
| | *Cytosol* | 33750 | 71957 | 0.77 | - | - |
| | *Mitochondrion* | 7541 | 17509 | 0.75 | - | - |
| | *Nucleus* | 29789 | 57722 | 0.79 | - | - |
| | *Extracellular* | 11672 | 35729 | 0.80 | - | - |
| | *Secretory Pathway* | 13460 | 36444 | 0.77 | - | - |
| | *Membrane* | 27013 | 61050 | 0.76 | - | - |
| *D. melanogaster* | *All Localizations* | 21635 | 31886 | 0.55 | - | - |
| | *Cytosol* | 9192 | 16260 | 0.60 | - | - |
| | *Mitochondrion* | 1907 | 3598 | 0.57 | - | - |
| | *Nucleus* | 7344 | 12090 | 0.58 | - | - |
| | *Extracellular* | 2178 | 4714 | 0.58 | - | - |
| | *Secretory Pathway* | 4918 | 9359 | 0.54 | - | - |
| | *Membrane* | 6347 | 10607 | 0.60 | - | - |
| *C. elegans* | *All Localizations* | 20046 | 29281 | 0.73 | - | - |
| | *Cytosol* | 7713 | 13887 | 0.74 | - | - |
| | *Mitochondrion* | 1780 | 3521 | 0.72 | - | - |
| | *Nucleus* | 6245 | 10924 | 0.74 | - | - |
| | *Extracellular* | 1662 | 3325 | 0.70 | - | - |
| | *Secretory Pathway* | 4691 | 8895 | 0.72 | - | - |
| | *Membrane* | 7190 | 10579 | 0.72 | - | - |
| *S. cerevisiae* | *All Localizations* | 6480 | 11423 | 0.81 | - | - |
| | *Cytosol* | 3467 | 7221 | 0.81 | - | - |
| | *Mitochondrion* | 1513 | 3124 | 0.76 | - | - |
| | *Nucleus* | 2926 | 5681 | 0.81 | - | - |
| | *Extracellular* | 282 | 762 | 0.76 | - | - |
| | *Secretory Pathway* | 999 | 2477 | 0.81 | - | - |
| | *Membrane* | 2237 | 4562 | 0.82 | - | - |

Supplementary Table S3. **Comparison of the ComPPI content to the input databases, and the effects of our filtering algorithms and manual validation steps.**

The table shows the number of interaction and localization entries in the source databases and the number of them loaded into ComPPI. Different input sources are connected to the ComPPI database structure using source-specific interfaces. During autoloading or manual validation steps (see Figure 1 for more details in the main text) we filtered out those interactions or localizations from the source databases, that (1) did not the requirements of ComPPI (e.g. genetic interactions in BioGRID or localizations in PA-GOSUB with a confidence level below 95%), (2) contained errors in their data structure (e.g. entries with inconsequent nomenclature), or those that (3) turned out to be biologically unlikely during our manual review process. We also mapped the different subcellular localization naming conventions to GO (8) cellular component terms for standardization purposes (Supplementary Figure S2). The source databases have different protein naming conventions, thus we had to map these protein names to the most reliable naming convention (visit the relevant Help page for more details: http://comppi.linkgroup.hu/help/naming_conventions). Due to the inconsistencies in protein naming conventions some protein names may be mapped to multiple other protein names. For instance, gene IDs could be mapped to several protein IDs, which phenomenon is based on real biological processes, such as alternative splicing. This may result in more protein names associated with a given source than the number of proteins taken from the original source. There are some other cases, where protein names could not be mapped to the strongest protein naming convention. In these cases we dropped the entry, so it was not incorporated into the database. Another important point is that we developed an algorithm in order to export the predefined datasets from the ComPPI database structure. The export module also went through rigorous manual revision in order to ensure that there are exact matches between the source data and the output data from ComPPI (see Supplementary Figure S3 and Table S2 for more details about our output data). Relevant information of the efficiency of manual curation could be gained in those cases, where source databases, such as MatrixDB and HPRD use matching protein name conventions and have a consequent data structure. Taking these facts together this table shows the summarized effect of filtering due to the manual curation protocols, the filtering due to our special requirements of incorporated data, and the effect of the protein name mapping.

| \textit{Saccharomyces cerevisiae} | | | | | | |
|---|---|---|---|---|---|---|
| **Protein-protein Interaction Databases** | | | | | | |
| Source Database | **BioGRID** | **CCSB** | **DiP** | **IntAct** | **MINT** | |
| Number of interactions loaded into ComPPI | 82358 | 3328 | 22970 | 77216 | 24945 | |
| Number of all the interactions in the source database | 340723 | 2930 | 22735 | 124582 | 48628 | |
| **Subcellular Localization Databases** | | | | | | |
| Source Database | **eSLDB** | **GeneOntology** | **OrganelleDB** | **PA-GOSUB** | | |
| Number of localizations loaded into ComPPI | 8424 | 12230 | 7568 | 3421 | | |
| Number of all the localizations in the source database | 8581 | 63338 | 8237 | 273944 | | |
| \textit{Caenorhabditis elegans} | | | | | | |
| **Protein-protein Interaction Databases** | | | | | | |
| Source Database | **BioGRID** | **CCSB** | **DiP** | **IntAct** | **MINT** | |
| Number of interactions loaded into ComPPI | 15735 | 9050 | 3942 | 11552 | 5358 | |
| Number of all the interactions in the source database | 8464 | 3864 | 4107 | 20342 | 7400 | |
| **Subcellular Localization Databases** | | | | | | |
| Source Database | **eSLDB** | **GeneOntology** | **OrganelleDB** | **PA-GOSUB** | | |
| Number of localizations loaded into ComPPI | 23465 | 13589 | 544 | 11946 | | |
| Number of all the localizations in the source database | 33336 | 56511 | 551 | 974028 | | |
| \textit{Drosophila melanogaster} | | | | | | |
| **Protein-protein Interaction Databases** | | | | | | |
| Source Database | **BioGRID** | **DiP** | **DroID** | **IntAct** | **MINT** | |
| Number of interactions loaded into ComPPI | 100517 | 24375 | 198533 | 26161 | 22413 | |
| Number of all the interactions in the source database | 47573 | 23154 | 96023 | 30183 | 23548 | |
| **Subcellular Localization Databases** | | | | | | |
| Source Database | **eSLDB** | **GeneOntology** | **OrganelleDB** | **PA-GOSUB** | | |
| Number of localizations loaded into ComPPI | 22907 | 20971 | 3855 | 11070 | | |
| Number of all the localizations in the source database | 20815 | 112652 | 3816 | 711788 | | |

| *Homo sapiens* | | | | | | | | |
|---|---|---|---|---|---|---|---|---|
| **Protein-protein Interaction Databases** | | | | | | | | |
| Source Database | **BioGRID** | **CCSB** | **DiP** | **HPRD** | **IntAct** | **MatrixDB** | **MINT** | **MIPS** |
| Number of interactions loaded into ComPPI | 363880 | 3733 | 6663 | 37180 | 62070 | 148 | 23206 | 421 |
| Number of all the interactions in the source database | 230603 | 3881 | 5951 | 39240 | 101128 | 1064 | 33259 | 1814 |
| **Subcellular Localization Databases** | | | | | | | | |
| Source Database | **eSLDB** | **GeneOntology** | **HumanProteinAtlas** | **HumanProteinpedia** | **LOCATE** | **MatrixDB** | **Organelle** | **PA-GOSUB** |
| Number of localizations loaded into ComPPI | 67487 | 83548 | 37509 | 2820 | 18822 | 9975 | 4886 | 21641 |
| Number of all the localizations in the source database | 81988 | 403734 | 9122 | 2900 | 18724 | 9975 | 4955 | 1566180 |

Supplementary Table S4. **Mapping of high resolution subcellular localization data into major cellular components.**

Subcellular localization data come from several sources with different resolution. Therefore, the integration of high resolution data into major cellular components with low resolution is needed. The high resolution localization data were mapped manually to the possibly largest and most accurate subcellular localizations based on the hierarchical branches (parent and children branches) of the localization tree (Supplementary Figure S2). One or more parent branches were associated with one of the 6 major subcellular components (cytosol, nucleus, mitochondrion, secretory-pathway, membrane and extracellular) as shown on the table. Thus, using the united, hierarchical localization tree we gained low resolution major cellular components, in which there is an unambiguous route in the tree to one major subcellular component. If a given GO term belongs to an included branch, but it is located in another major cellular component, then this GO term will be excluded during the mapping. Currently the localization tree contains 1,644 GO cellular component terms. The number of GO terms belonging to a given major cellular component is also shown in the last column of the table. The mapping table of major cellular components is available online here:

http://bificomp2.sote.hu:22422/comppi/files/85e7056adb541d5a18c60792457986c71a3a0ab0/databases/loctree/largelocs.yml.

| MAJOR CELLULAR COMPONENT | GO TERMS OF THE INCLUDED LOCALIZATION TREE BRANCHES | GO TERMS OF THE EXCLUDED LOCALIZATION TREE BRANCHES | NUMBER OF ALL INCLUDED GO TERMS |
|---|---|---|---|
| CYTOSOL | GO:0032994, GO:0002189, GO:0031501, GO:0034464, GO:0032992, GO:0097057, GO:0008247, GO:0009341, GO:0000151, GO:0005942, GO:0070864, GO:0043025, GO:0000502, GO:0035693, GO:0035859, GO:0000131, GO:0097169, GO:0072559, GO:0072557, GO:0044609, GO:0017122, GO:0033256, GO:0031074, GO:0030956, GO:0043226, GO:0042763, GO:0019008, GO:0008287, GO:0009349, GO:0030256, GO:0015627, GO:0012505, GO:0000267, GO:0032153, GO:0044297, GO:0045177, GO:0045178, GO:0016234, GO:0005930, GO:0005737, GO:0070725, GO:0005727, GO:0009368, GO:0000178, GO:0032144, GO:0032996, GO:0005956, GO:0033593, GO:0005953, GO:0005954, GO:0005952, GO:0030014, GO:0031431, GO:0009330, GO:0017101, GO:0031588, GO:0045259, GO:0000408, GO:0033588, GO:0005853, GO:0005850, GO:0005851, GO:0005852, GO:0016281, GO:0009318, GO:0034708, GO:0000307, GO:0017122, GO:0032045, GO:0035301, GO:0033256, GO:0008043, GO:0048269, GO:0070438, GO:0072558, GO:0070419, GO:0045252, GO:0043626, GO:0032299, GO:0071141, GO:0032797, GO:0009337, GO:0031931, GO:0031932, GO:0072669, GO:0031371, GO:0034657, GO:0033202, GO:0031074, GO:0034518, GO:0044446, GO:0031094, GO:0009536, GO:0005773, GO:0042579, GO:0045170, GO:0032421, GO:0005818, GO:0009295, GO:0043292, GO:0005856, GO:0043227, GO:0043228, GO:0031090, GO:0031974 | GO:0005623, GO:0044464, GO:0043231, GO:0043232, GO:0043229, GO:0044424, GO:0005622 | **535 / 1644** |
| NUCLEUS | GO:0005634, GO:0005694, GO:0005667, GO:0035649, GO:0000974, GO:0002193, GO:0097196, GO:0000441 | - | **353 / 1644** |
| MITOCHONDRION | GO:0005739, GO:0070069, GO:0016507, GO:0097136 | - | **78 / 1644** |
| SECRETORY-PATHWAY | GO:secretory_pathway, GO:0005783, GO:0005793, GO:0005794, GO:0005801, GO:0005802, GO:0005768, GO:0031410, GO:0042175, GO:0042175, GO:0031982 | - | **185 / 1644** |
| MEMBRANE | GO:0038037, GO:0070195, GO:0002116, GO:0035692, GO:0005892, GO:0048179, GO:0008328, GO:0042101, GO:0008305, GO:0019814, GO:0002133, GO:0009986, GO:0042597, GO:0071944, GO:0042995, GO:0030428, GO:0070938, GO:0030496, GO:0002139, GO:0031252, GO:0001917, GO:0043209, GO:0005933, GO:0060187, GO:0019867, GO:0044425, GO:0005886, GO:0030054, GO:0045202 | GO:0016020 | **367 / 1644** |
| EXTRACELLULAR | GO:0005576, GO:0031012, GO:0043033, GO:0031838, GO:0097179 | - | **126 / 1644** |

Supplementary Table S5. **Importance of the removal of localization-based biologically unlikely interactions.**

The table demonstrates the importance of the removal of biologically unlikely interactions based on their localization data. We compared the degree and the betweeness centrality (BC) of the proteins in the integrated protein-protein interaction (PPI) dataset without subcellular localizations and to the compartmentalized interactome, where the proteins have at least one common subcellular localization. The proteins were ranked by the size of the alternations in the degree (Δ Degree (%)) and BC measures, and only the proteins with UniProt Swiss-Prot (28) name were kept (15,258 proteins out of 19,386) for more reliable further analysis. The table shows the first 20 proteins, ordered by the degree difference between the compartmentalized dataset and the integrated PPI dataset (Δ Degree). The subcellular localization of the proteins are also displayed, these can be: (i) subcellular localizations after the compartmentalization (Compartmentalized Localization), (ii) localizations that were represented in the first-neighbours of the interactome with less than 5% ratio (Filtered Localizations), (iii) the predicted localizations, where more than 95% of the first-neighbours are localized in the given compartment (Possible Localizations), and (iv) the localizations that were represented among the first-neighbours with a range between 5 and 95% (Uncertain Localizations). The table gives information about how many proteins have no localization data among the first-neighbours, and gives an average Interaction Score (IS) for the interactions before and after filtering. Our application example is the Crotonase (Enoyl-CoA hydratase), which is the first in the list (highlighted in yellow). See Figure 4 in the main text for more details.

| NUMBER | UNPROT SWISS-PROT ID | RECOMMENDED PROTEIN NAME | DEGREE | | BETWEENESS CENTRALITY | | Δ DEGREE | Δ DEGREE (%) |
|---|---|---|---|---|---|---|---|---|
| | | | INTEGRATED PPI DATASET | COMPARTMENTALIZED PPI DATASET | INTEGRATED PPI DATASET | COMPARTMENTALIZED PPI DATASET | | |
| 1 | P30084 | Enoyl-CoA hydratase, mitochondrial | 71 | 8 | 4855.94 | 107.43 | 63 | 11.27% |
| 2 | Q96IL0 | Apoptogenic protein 1, mitochondrial | 57 | 7 | 7072.36 | 9.97 | 50 | 12.28% |
| 3 | Q96EY7 | Pentatricopeptide repeat domain-containing protein 3, mitochondrial | 47 | 7 | 8080.17 | 69.29 | 40 | 14.89% |
| 4 | Q9Y3A4 | Ribosomal RNA-processing protein 7 homolog A | 41 | 3 | 409.31 | 2.02 | 38 | 7.32% |
| 5 | Q8IZ73 | RNA pseudouridylate synthase domain-containing protein 2 | 27 | 2 | 1911.21 | 0.37 | 25 | 7.41% |
| 6 | O14874 | [3-methyl-2-oxobutanoate dehydrogenase [lipoamide]] kinase, mitochondrial | 26 | 3 | 15548.09 | 5.00 | 23 | 11.54% |
| 7 | Q53H82 | Beta-lactamase-like protein 2 | 17 | 1 | 3093.07 | 0.00 | 16 | 5.88% |
| 8 | Q99551 | Transcription termination factor, mitochondrial | 15 | 1 | 176.46 | 0.00 | 14 | 6.67% |
| 9 | Q9BSE5 | Agmatinase, mitochondrial | 12 | 1 | 350.66 | 0.00 | 11 | 8.33% |
| 10 | P23378 | Glycine dehydrogenase (decarboxylating), mitochondrial | 10 | 1 | 1625.93 | 0.00 | 9 | 10.00% |
| 11 | Q9P0P8 | Uncharacterized protein C6orf203 | 9 | 1 | 645.26 | 0.00 | 8 | 11.11% |
| 12 | P54868 | Hydroxymethylglutaryl-CoA synthase, mitochondrial | 8 | 1 | 30.12 | 0.00 | 7 | 12.50% |
| 13 | Q9Y680 | Peptidyl-prolyl cis-trans isomerase FKBP7 | 8 | 1 | 174.14 | 0.00 | 7 | 12.50% |
| 14 | O95377 | Gap junction beta-5 protein | 7 | 1 | 305.81 | 0.00 | 6 | 14.29% |
| 15 | P54317 | Pancreatic lipase-related protein 2 | 7 | 1 | 4111.91 | 0.00 | 6 | 14.29% |
| 16 | Q08E93 | Protein FAM27E3 | 7 | 1 | 249.87 | 0.00 | 6 | 14.29% |
| 17 | Q5T7N7 | Putative protein FAM27E1 | 7 | 1 | 249.87 | 0.00 | 6 | 14.29% |
| 18 | Q6P4F2 | Adrenodoxin-like protein, mitochondrial | 7 | 1 | 4496.69 | 0.00 | 6 | 14.29% |
| 19 | Q9BV35 | Calcium-binding mitochondrial carrier protein SCaMC-3 | 7 | 1 | 86.76 | 0.00 | 6 | 14.29% |
| 20 | Q9Y234 | Lipoyltransferase 1, mitochondrial | 7 | 1 | 0 | 0.00 | 6 | 14.29% |

| NUMBER | COMPARTMENTALIZED LOCALIZATION | FILTERED LOCALIZATIONS (<5%) | POSSIBLE LOCALIZATIONS (>95%) | UNCERTAIN LOCALIZATIONS (5%<X<95%) | NUMBER OF LOCALIZATIONS | AVERAGE INTERACTION SCORE BEFORE FILTERING | AVERAGE INTERACTION SCORE AFTER FILTERING |
|---|---|---|---|---|---|---|---|
| 1 | Mitochondrion (8/71) | - | - | Cytosol (52/71), Nucleus (33/71), Membrane (27/71), Extracellular (21/71), Secretory-pathway (17/71) | 56/71 | 0.075 | 0.665 |
| 2 | Mitochondrion (7/57) | - | - | Cytosol (38/57), Nucleus (28/57), Membrane (19/57), Secretory-pathway (18/57), Extracellular (14/57) | 48/62 | 0.057 | 0.463 |
| 3 | Mitochondrion (7/47) | - | - | Nucleus (34/47), Cytosol (31/47), Membrane (16/47), Extracellular (14/47), Secretory-pathway (13/47) | 41/47 | 0.137 | 0.918 |
| 4 | Mitochondrion (3/41) | Secretory-pathway (2/41) | - | Nucleus (34/41), Cytosol (25/41), Membrane (12/41), Extracellular (5/41) | 36/41 | 0.033 | 0.457 |
| 5 | Mitochondrion (2/27) | - | - | Cytosol (23/27), Nucleus (12/27), Membrane (8/27), Secretory-Pathway (7/27), Extracellular (5/27) | 24/27 | 0.02 | 0.273 |
| 6 | Mitochondrion (3/26) | - | - | Cytosol (18/26), Nucleus (17/26), Membrane (8/26), Secretory-pathway (6/26), Extracellular (5/26) | 23/26 | 0.096 | 0.833 |
| 7 | Mitochondrion (1/17) | - | - | Nucleus (13/17), Cytosol (12/17), Membrane (5/17), Extracellular (2/17), Secretory-pathway (2/17) | 16/17 | 0.037 | 0.632 |
| 8 | Mitochondrion (1/15) | - | - | Cytosol (11/15), Nucleus (10/15), Membrane (7/15), Extracellular (5/15), Secretory-pathway (4/15) | 14/15 | 0.067 | 0.999 |
| 9 | Mitochondrion (1/12) | - | - | Nucleus (8/12), Cytosol (4/12), Secretory-pathway (1/12), Membrane (1/12), Extracellular (1/12) | 10/12 | 0.077 | 0.92 |
| 10 | Mitochondrion (1/10) | - | - | Nucleus (7/10), Cytosol (5/10), Secretory-pathway (2/10), Membrane (2/10), Extracellular (2/10) | 8/10 | 0.094 | 0.936 |
| 11 | Mitochondrion (1/9) | - | - | Cytosol (6/9), Nucleus (5/9), Secretory-pathway (4/9), Membrane (4/9), Extracellular (1/9) | 7/9 | 0.104 | 0.939 |
| 12 | Mitochondrion (1/8) | - | - | Cytosol (6/8), Secretory-pathway (6/8), Membrane (5/8), Nucleus (4/8), Extracellular (3/8) | 6/8 | 0.037 | 0.299 |
| 13 | Secretory-pathway (1/8) | Mitochondrion (0/8), Membrane (0/8), Extracellular (0/8) | - | Cytosol (5/8), Nucleus (5/8) | 6/8 | 0.116 | 0.928 |
| 14 | Membrane (1/7), Extracellular (1/7) | Mitochondrion (0/7), Secretory-pathway (0/7) | - | Cytosol (4/7), Nucleus (4/7) | 7/7 | 0.139 | 0.973 |
| 15 | Secretory-pathway (1/7), Extracellular (1/7), Membrane (1/7) | - | - | Cytosol (1/7), Mitochondrion (1/7), Nucleus (1/7) | 3/7 | 0.141 | 0.988 |
| 16 | Mitochondrion (1/7) | - | - | Cytosol (4/7), Nucleus (4/7), Extracellular (3/7), Membrane (3/7), Secretory-pathway (2/7) | 7/7 | 0.08 | 0.56 |
| 17 | Mitochondrion (1/7) | - | - | Cytosol (4/7), Nucleus (4/7), Extracellular (3/7), Membrane (3/7), Secretory-pathway (2/7) | 7/7 | 0.08 | 0.56 |
| 18 | Mitochondrion (1/7) | - | - | Cytosol (4/7), Membrane (3/7), Secretory-pathway (3/7), Nucleus (2/7), Extracellular (1/7) | 5/7 | 0.134 | 0.937 |
| 19 | Membrane (1/7) | Mitochondrion (0/7), Extracellular (0/7) | - | Nucleus (5/7), Cytosol (3/7), Secretory-pathway (2/7) | 5/7 | 0.041 | 0.287 |
| 20 | Mitochondrion (1/7) | | | Cytosol (5/7), Membrane (4/7), Secretory-pathway (4/7), Nucleus (3/7), Extracellular (2/7) | 5/7 | 0.099 | 0.691 |

Supplementary Table S6. **The results of the Gene Ontology biological process enrichment analysis of the two example proteins, crotonase and MPS1.**

The table shows the results of the Gene Ontology (8) biological process term enrichment analysis using BiNGO (17). Our two application examples are crotonase (see Figure 4 in the main text for more details) and MPS1 (see Supplementary Figure S7 for more details) to demonstrate how ComPPI is useful in the filtering of biologically unlikely interactions and prediction in the new protein properties and functions. The upper part of the table shows the first 15 biological process terms in order of their corrected significance from the analysis of the mitochondrial subset of the first-neighbours of crotonase. Importantly, the most significant biological process is the 'anti-apoptosis' (highlighted in yellow), while related terms are also highly represented in the list. Known functions of crotonase, such as 'positive regulation of lipopolysaccharide-mediated signalling pathway' (highlighted in green) or 'catabolic process' are underrepresented in the list, which implicates the importance of a putative role of crotonase in the inhibition of apoptosis. The bottom part of the table shows the first 15 biological process terms from the results of the GO enrichment analysis of the nuclear subset of the first-neighbours of MPS1. The analysis showed that beside the known functions, such as 'cell cycle process' (highlighted in green) and related terms, biological processes in connection with the proposed functions in nuclear assembly also occurred significantly, such as 'cellular component organization' (highlighted in yellow) and related terms. The two example proteins show the usefulness of ComPPI to predict new biological functions of various proteins, and to understand their role in cellular functions better.

| | | GO BIOLOGICAL PROCESS ENRICHMENT ANALYSIS - CROTONASE | | | |
|---|---|---|---|---|---|
| **NUMBER** | **GO-ID** | **DESCRIPTION** | **CORRECTED P-VALUE** | **CLUSTER FREQUENCY** | **GENES** |
| 1 | 6916 | anti-apoptosis | 2.26E-03 | 4/9 44.4% | P40337 \| Q9Y4K3 \| P38646 \| P63104 |
| 2 | 43066 | negative regulation of apoptosis | 6.17E-03 | 4/9 44.4% | P40337 \| Q9Y4K3 \| P38646 \| P63104 |
| 3 | 43069 | negative regulation of programmed cell death | 6.17E-03 | 4/9 44.4% | P40337 \| Q9Y4K3 \| P38646 \| P63104 |
| 4 | 60548 | negative regulation of cell death | 6.17E-03 | 4/9 44.4% | P40337 \| Q9Y4K3 \| P38646 \| P63104 |
| 5 | 42981 | regulation of apoptosis | 6.17E-03 | 5/9 55.5% | P21980 \| P40337 \| Q9Y4K3 \| P38646 \| P63104 |
| 6 | 43067 | regulation of programmed cell death | 6.17E-03 | 5/9 55.5% | P21980 \| P40337 \| Q9Y4K3 \| P38646 \| P63104 |
| 7 | 10941 | regulation of cell death | 6.17E-03 | 5/9 55.5% | P21980 \| P40337 \| Q9Y4K3 \| P38646 \| P63104 |
| 8 | 48661 | positive regulation of smooth muscle cell proliferation | 9.22E-03 | 2/9 22.2% | P21980 \| Q9Y4K3 |
| 9 | 48660 | regulation of smooth muscle cell proliferation | 1.90E-02 | 2/9 22.2% | P21980 \| Q9Y4K3 |
| 10 | 2274 | myeloid leukocyte activation | 1.93E-02 | 2/9 22.2% | Q9Y4K3 \| P63104 |
| 11 | 35148 | tube formation | 1.93E-02 | 2/9 22.2% | P21980 \| Q9Y4K3 |
| 12 | 48771 | tissue remodeling | 1.93E-02 | 2/9 22.2% | P21980 \| Q9Y4K3 |
| 13 | 31666 | positive regulation of lipopolysaccharide-mediated signalling pathway | 1.93E-02 | 1/9 11.1% | Q9Y4K3 |
| 14 | 2441 | histamine secretion involved in inflammatory response | 1.93E-02 | 1/9 11.1% | P63104 |
| 15 | 290 | deadenylation-dependent decapping of nuclear-transcribed mRNA | 1.93E-02 | 1/9 11.1% | Q96C86 |
| | | GO BIOLOGICAL PROCESS ENRICHMENT ANALYSIS - MPS1 | | | |
| **NUMBER** | **GO-ID** | **DESCRIPTION** | **CORRECTED P-VALUE** | **CLUSTER FREQUENCY** | **GENES** |
| 1 | 22402 | cell cycle process | 8.91E-09 | 11/19 57.8% | P04637 \| P07437 \| Q53HL2 \| Q13042 \| Q8NG31 \| Q12834 \| P30260 \| P24941 \| P05067 \| Q9H4B7 \| Q53GA5 |
| 2 | 7049 | cell cycle | 8.91E-09 | 12/19 63.1% | P04637 \| P07437 \| Q16659 \| Q53HL2 \| Q13042 \| Q8NG31 \| Q12834 \| P30260 \| P24941 \| P05067 \| Q9H4B7 \| Q53GA5 |
| 3 | 22403 | cell cycle phase | 3.05E-07 | 9/19 47.3% | P07437 \| Q53HL2 \| Q13042 \| Q8NG31 \| Q12834 \| P30260 \| P24941 \| P05067 \| Q9H4B7 |
| 4 | 279 | M phase | 1.06E-06 | 8/19 42.1% | P07437 \| Q53HL2 \| Q13042 \| Q8NG31 \| Q12834 \| P30260 \| P24941 \| Q9H4B7 |
| 5 | 280 | nuclear division | 1.06E-06 | 7/19 36.8% | P07437 \| Q53HL2 \| Q13042 \| Q8NG31 \| Q12834 \| P30260 \| P24941 |
| 6 | 7067 | Mitosis | 1.06E-06 | 7/19 36.8% | P07437 \| Q53HL2 \| Q13042 \| Q8NG31 \| Q12834 \| P30260 \| P24941 |
| 7 | 278 | mitotic cell cycle | 1.06E-06 | 8/19 42.1% | P07437 \| Q53HL2 \| Q13042 \| Q8NG31 \| Q12834 \| P30260 \| P24941 \| P05067 |
| 8 | 87 | M phase of mitotic cell cycle | 1.06E-06 | 7/19 36.8% | P07437 \| Q53HL2 \| Q13042 \| Q8NG31 \| Q12834 \| P30260 \| P24941 |
| 9 | 48285 | organelle fission | 1.06E-06 | 7/19 36.8% | P07437 \| Q53HL2 \| Q13042 \| Q8NG31 \| Q12834 \| P30260 \| P24941 |
| 10 | 32270 | positive regulation of cellular protein metabolic process | 3.56E-06 | 7/19 36.8% | P07437 \| Q53HL2 \| Q13042 \| Q8NG31 \| Q12834 \| P30260 \| P24941 |
| 11 | 51247 | positive regulation of protein metabolic process | 4.77E-06 | 7/19 36.8% | P04637 \| Q13042 \| Q12834 \| P30260 \| P55072 \| Q96IF9 \| Q53GA5 |
| 12 | 34976 | response to endoplasmic reticulum stress | 6.74E-06 | 4/19 21.0% | P04637 \| P55072 \| Q96IF9 \| Q53GA5 |
| 13 | 6984 | ER-nucleus signaling pathway | 7.31E-06 | 4/19 21.0% | P04637 \| P55072 \| Q96IF9 \| Q53GA5 |
| 14 | 16043 | cellular component organization | 7.31E-06 | 14/19 73.6% | P07437 \| Q53HL2 \| Q13042 \| Q8NG31 \| Q12834 \| P05067 \| P24941 \| Q53GA5 \| P04637 \| P30260 \| P61964 \| Q96IF9 \| P55072 \| Q9H4B7 |
| 15 | 6996 | organelle organization | 1.02E-05 | 11/19 57.8% | P04637 \| P07437 \| Q53HL2 \| Q13042 \| Q8NG31 \| Q12834 \| P30260 \| P24941 \| P61964 \| Q9H4B7 \| Q53GA5 |